\documentclass[aip,reprint]{revtex4-1}
\usepackage{amsmath}

\usepackage{xcolor}
\usepackage{graphicx}

\def\br{{\bf r}}
\def\bp{{\bf r}'}
\def\dd{{\rm d}}
\draft 

\begin{document}

\title{Decomposition and embedding in the stochastic $GW$ self-energy}

\author{Mariya Romanova}
\author{Vojt\v{e}ch Vl\v{c}ek}
\email{vlcek@ucsb.edu}
\affiliation{Department of Chemistry and Biochemistry, University of California, Santa Barbara, CA 93106-9510, U.S.A.}

\begin{abstract}
We present two new developments for computing excited state energies within the $GW$ approximation. First, calculations of the Green's function and the screened Coulomb interaction are decomposed into two parts: one is deterministic while the other relies on stochastic sampling. Second, this separation allows constructing a subspace self-energy, which contains dynamic correlation from only a particular (spatial or energetic) region of interest. The methodology is exemplified on large-scale simulations of nitrogen-vacancy states in a periodic hBN monolayer and hBN-graphene heterostructure. We demonstrate that the deterministic embedding of strongly localized states significantly reduces statistical errors, and the computational cost decreases by more than an order of magnitude. The computed subspace self-energy unveils how interfacial couplings affect electronic correlations and identifies contributions to excited-state lifetimes. While the embedding is necessary for the proper treatment of impurity states, the decomposition yields new physical insight into quantum phenomena in heterogeneous systems.

\end{abstract}

\maketitle

\section{Introduction}

First-principles treatment of electron excitation energies is crucial for guiding the development of new materials with tailored optoelectronic properties. Localized quantum states became the focal point for condensed systems:\cite{Yu_2017,Grosso2017,Yankowitz2018,Cao2018,Zondiner2020,Turiansky2020,Gottscholl2020} due to their long-range electronic correlation, the localized excitations exhibit tunability by interfacial phenomena.\cite{Wang_2015,Qiu2017,Tartakovskii2020,Yuan2020} Besides predicting experimental observables, theoretical investigations thus help to elucidate the interplay of the electron-electron interactions and the role of the environment.

The excited electrons and holes near the Fermi level are conveniently described by the quasiparticle (QP) picture: the charge carriers are characterized by renormalized interactions and a finite lifetime limited by energy dissipation, which governs the deexcitation mechanism. Quantitative predictions of QPs necessitate the inclusion of non-local many-body interactions. 

The prevalent route to describe QPs in condensed systems employs the Green's function formalism.\cite{martin2016interacting,FetterWalecka} The excitation energy and its lifetime are inferred from the QP dynamics. The many-body effects are represented by the non-local and dynamical self-energy, $\Sigma$. In practice, $\Sigma$ is approximated by selected classes of interactions, forming a hierarchy of systematically improvable methods.\cite{martin2016interacting} The formalism also allows constructing the self-energy for distinct states with a different form of $\Sigma$.

Here we neglect the vibrational effects, and the induced density fluctuations dominate the response of the system to the excitation. The perturbation expansion is conveniently based on the screened Coulomb interaction, $W$, which explicitly incorporates the system's polarizability. The neglect of the induced quantum fluctuations (while accounting for the induced density) leads to the popular $GW$ method.\cite{Hedin1965,Hybertsen_1986,Aryasetiawan1998,martin2016interacting} This approach predicts the QP gap, ionization potentials, and electron affinities in excellent agreement with experiments.\cite{Hedin1965, Aryasetiawan1998} Within the $GW$ approximation, $\Sigma$ is a product of $W$ and the Green's function, $G$, in the time domain. 

Conventional implementations of the $GW$ self-energy scale as $O(N^4)$ with the system size, albeit with a small prefactor,\cite{Govoni2015} and it is usually limited to few-electron systems. Stochastic treatment recasts the self-energy as a statistical estimator and employs sampling of electronic wavefunctions combined with the  decomposition of operators.\cite{neuhauser2014breaking, Vlcek2018swift,vlcek2017stochastic} In practice, this formulation decreases the computational time significantly and leads to a linear scaling algorithm.\cite{neuhauser2014breaking,vlcek2017stochastic,Vlcek2018swift}

The stochastic approach uses real-space random vectors that sample the Hilbert space of the electronic Hamiltonian. The statistical error in $\Sigma$ is governed by the number of vectors, which are used for the decomposition of $G$ and the evaluation of $W$. While extended systems with delocalized states are treated efficiently, the statistical fluctuations are more significant for localized orbitals.\cite{Brooks_2020} This translates to an increased computational cost for defects, impurities, and molecules.\cite{Brooks_2020,vlcek2017stochastic} Further, too high fluctuations potentially hinder proper convergence and may result in sampling bias.

Here we overcome this difficulty by constructing a hybrid deterministic-stochastic approach. We show how to efficiently decompose $G$  and $W$ in real-time and embed the strongly localized states. In this formalism,  the problematic orbitals are treated explicitly without relying on their sampling by random vectors. A similar embedding scheme was so far employed only in static ground-state calculations.\cite{Li_2019} The decomposition is general and can be used to sample arbitrary excitations. Naturally, it is especially well suited for spatially or energetically isolated electronic states. Further, we employ the decomposition technique to compute $\Sigma$ stochastically from an arbitrarily large subspace of interest. We show the self-energy separation disentangles correlation contributions from different spatial regions of the system. 

After deriving the formalism of the self-energy embedding and decomposition, we illustrate our methods numerically for nitrogen vacancy in large periodic cells of hBN. This system represents a realistic  simulation of a prototypical single-photon emitter.\cite{Tran_2016,Grosso2017,Exarhos_2017,Li_2019,Mendelson_2019} First, we demonstrate the reduction of the stochastic error for the defect states in the hBN monolayer. Next, we study the interaction of the defect in the hBN-graphene heterostructure. 

\section{Computing quasiparticle energies}

\subsection{Self-energy and common approximations}

The Green's function ($G$),  a time-ordered correlator of creation and annihilation field operators, describes the dynamics of an individual quasiparticle. The poles of $G$ fully determine the single-particle excitations (as well as many other properties). 

Solving directly for G  is often technically challenging (or outright impossible).  Alternatively, the Green's function is often sought via a perturbative expansion of the electron-electron interactions on top of a propagator of non-interacting particles (i.e., the non-interacting Green's function, $G_0$). The two quantities are related via the Dyson equation $G^{-1} = G_0^{-1} - \Sigma$, where $\Sigma$ is a self-energy accounting, in principle, for all the many-body effects absent in $G_0$.

Calculations usually employ only a truncated expansion of the self-energy. Despite ongoing developments,\cite{ Maggio2017, Hellgren2018, Vlcek2019vertex} the most common approach is limited to the popular $GW$ approximation to the self-energy, which is composed of the non-local exchange ($\Sigma_X$) and polarization  ($\Sigma_P$) terms. In the time-domain, the latter operator is expressed as:
\begin{equation}\label{eq:sigma_GW}
\Sigma_P(\br,\br',t) = iG(\br,\br',t)W_P(\br,\br',t^+),
\end{equation}
where $W_P$ is the time-ordered polarization potential due to the  time-dependent induced charge density \cite{Vlcek2018swift}. The potential is conveniently expressed using the reducible polarizability $\chi$ as
\begin{equation}
\begin{split}
 W_P(\br,&\br',t) = \\ 
 &\int\int \nu(\br,\br'') \chi(\br''\br''',t) \nu(\br''',\br')d\br''d\br''' 
\end{split}
\label{polarizability}
\end{equation}

Evaluating the action of $W_P$ on individual states is the practical bottleneck of the $GW$ approach. Hence, despite Eq.~\ref{eq:sigma_GW} requires a self-consistent solution, it is commonly computed only as a ``one-shot'' correction. In practice, Eq.~\ref{eq:sigma_GW} thus contains only $G_0$ and $W_0$. Both quantities are constructed from the mean-field Hamiltonian, $H_0$, comprising one-body terms and local Hartree, ionic, and exchange-correlation potentials. For $W_P$, it is common to employ the random phase approximation (RPA). Beyond RPA approaches are more expensive and, in general, do not improve the QP energies unless higher-order (vertex) terms are included in $\Sigma$.\cite{Lewis2019,Vlcek2019vertex}

Within this one-shot framework, the QP energies become\cite{martin2016interacting}
\begin{equation}
    \varepsilon^{QP} = \varepsilon^{0} + \langle\phi| \Sigma_X + \Sigma_P(\omega = \varepsilon^{QP}) - v_{xc} |\phi\rangle,
\label{QP_energy}    
\end{equation}
where $\varepsilon^{0}$ are eigenvalues of $H_0$ and $v_{xc}$ is the  (approximate) exchange-correlation potential. In Eq.~\ref{QP_energy},  $\Sigma_P$ is in the frequency domain.

\subsection{The stochastic approach to the self-energy}
\label{sec:stoc}

The stochastic $G_0W_0$ method seeks the QP energy via random sampling of wavefunctions and decomposition of operators in the real-time domain.\cite{neuhauser2014breaking,vlcek2017stochastic,Vlcek2018swift} The expectation value of $\Sigma$ is expressed as a statistical estimator. The result is subject to fluctuations that decrease with the number of samples as $1/\sqrt{N}$.

In this formalism, the polarization self-energy expression is separable.\cite{neuhauser2014breaking,vlcek2017stochastic,Vlcek2018swift} Specifically, for a particular $H_0$ eigenstate $\phi$, the perturbative correction becomes:
\begin{align}
\langle\phi|\Sigma_P(t)|\phi\rangle &= \langle\phi|iG_0(t)W_P(t)|\phi\rangle \nonumber\\
&\simeq \frac{1}{N_{ \bar{\zeta}}} \sum_{\bar{\zeta}} \int \phi(\br)\zeta(\br,t)u_\zeta(\br,t) d^3\br,
\label{self-energy}
\end{align}
where $u_{\zeta}(r,t)$ is an induced charge density potential, and $\zeta$ is a random vector used for sampling of $G_0$ (discussed in detail in the next section). The state $\zeta$ at time $t$ is 
\begin{equation}\label{tevolve}
|\zeta(t)\rangle~\equiv~U_{0,t} P_\mu |{\zeta}\rangle,    
\end{equation}
where the $U_{0,t}$ is time evolution operator
\begin{equation}\label{u_tprop}
    U_{0,t} \equiv e^{-i H_0 t}.
\end{equation}
The projector $P_\mu$ selects the states above or below the chemical potential, $\mu$, depending on the sign of $t$. In practice, $P_\mu$ is directly related to the Fermi-Dirac operator.\cite{BaerNeuhauser2004,Gao_2015,Neuhauser_2016,Vlcek2018swift}  Since the Green's function is a time-ordered quantity, the vectors in the occupied and unoccupied subspace are propagated backward or forward in time and contribute selectively to the hole and particle non-interacting Green's functions.

The induced potential  $u(\br,t)$ represents the  time-ordered potential of the response to the charge addition or removal:
\begin{equation}
u(\br,t) = \int W_P(\br,\br',t) \bar{\zeta}(\br')\phi(\br')d^3\br',
\label{eff_potential}
\end{equation}
where $\bar \zeta$ spans the entire Hilbert space.\cite{neuhauser2014breaking,vlcek2017stochastic,Vlcek2018swift}

In practice, we compute $u$ from the retarded response potential, which is $\tilde u = \int \tilde W_P (\br,\br',t) \bar{\zeta}(\br')\phi(\br')d^3\br'$; the time-ordering step affects only the imaginary components of the Fourier transforms of $u$ and $\tilde u$.\cite{FetterWalecka,vlcek2017stochastic,Vlcek2018swift} 

The retarded response, $\tilde u$, is directly related to the time-evolved charge density $\delta n (\vec r,t) \equiv n(\br,t) - n(\vec r, t=0)$ induced by a perturbing potential $\delta v$:
\begin{align}\label{eqn:u}
&\tilde u(\br,t) = \nonumber\\
\int\int\int \nu(\br,\br'') &\chi(\br'',\br''',t) \delta v(\br''',\br')d\br'd\br''d\br''' \nonumber \\
&\equiv \int \nu(\br,\bp) \delta n(\bp,t) {\dd}\bp
\end{align}
where we define a perturbing potential
\begin{equation}\label{perturbing_pot}
    \delta v = \nu(\br,\br'){\bar\zeta}(\br')\phi(\br').
\end{equation}
Note that $\delta v$ is explicitly dependent on the state $\phi$ and the $\bar\zeta$ vector; the latter is part of the stochastically decomposed $G_0$ operator.

The stochastic formalism further reduces the cost of evaluating $\tilde u$. Instead of computing $\delta n(\vec r,t)$ by a sum over single-particle states,  we use another set of random vectors $\left\{ \eta \right\}$ confined to the occupied subspace. Time-dependent density $n(\br,t)$  thus becomes\cite{neuhauser2014breaking,vlcek2017stochastic,Vlcek2018swift,Gao_2015,Rabani2015,Neuhauser_2016} 
\begin{equation}
n(\br,t)=\lim_{N_\eta \to \infty} \frac{1}{N_{\eta}}\sum_{l}^{N_{\eta}}|\eta_l(\br,t)|^2,
\label{TDdensity}
\end{equation}
where $\eta_l$ is propagated in time using $U_{0,t}$, and $H_0$ in  Eq.~\ref{u_tprop} is implicitly time-dependent. As common,\cite{Hybertsen_1986,martin2016interacting}
we resort to the density functional theory (DFT) starting point. Since $H_0$ is therefore a functional of $n(\vec r,t)$, the same holds for the time evolution operator:
\begin{equation}\label{tpropeta_Un}
    \left| \eta(t)\right\rangle = U_{0,t} [n(t)] \left| \eta\right\rangle.
\end{equation}
Further, we employ RPA when computing $\tilde u$; this corresponds evolution within the time-dependent Hartree approximation.\cite{Baroni2001,BaerNeuhauser2004,Neuhauser2005}

Practical calculations use only a limited number of random states. Consequently, the time evolved density exhibits random fluctuations at each space-time point. To resolve the response to $\delta v$, we use a two-step propagation whose difference is the $\delta n$ that typically converges fast with $N_\eta$.\cite{neuhauser2014breaking,vlcek2017stochastic,Vlcek2018swift}

\section{Embedded Deterministic Subspace}\label{sec:embed_theory}

The stochastic vectors $\left\{\zeta\right\}$ and $\left\{\eta\right\}$, introduced in the preceding section, are constructed on a real-space grid and sample the occupied (or unoccupied) states. The number of these vectors ($N_\zeta$ and $N_\eta$) is increased so that the statistical errors are below a predefined threshold. Further, the underlying assumption is that $\left\{\zeta\right\}$ and $\left\{\eta\right\}$ sample the Hilbert space uniformly.

Here, we present a stochastic approach restricted only to a subset of states, while selected orbitals, $\left\{\phi\right\}$, are treated explicitly and constitute an embedded subspace. We denote this set as the $\{\phi\}$-subspace. In the context of the $G_0W_0$ approximation, we use the hybrid approach for (i) the Green's function, (ii) the induced potential, or (iii) both $G_0$ and $u$ simultaneously. In the following, we present each case separately.

\paragraph{First,} the non-interacting Green's function is decomposed into two parts (omitting for brevity the space-time coordinates):
\begin{equation}\label{G0decomposed}
G_0 \equiv \tilde G_0 + G^\phi_0,    
\end{equation}
where $G^\phi_0$ is the Green's function of the constructed explicitly from $\left\{\phi\right\}$ as
\begin{equation}
     G^\phi_0(\br,\bp,t) = \sum_{j\in\{\phi\}} (-1)^{\theta(t)}\,i\phi_j(\br)\phi^*_j(\bp)e^{-i\varepsilon_j t} 
\end{equation}
where $\theta$ is the Heaviside step function responsible for the time-ordering (corresponding to particle and hole contributions to $G_0$). The complementary part, $\tilde G_0$, is sampled with random states as in Eq.~\ref{self-energy}.

Unlike in the fully stochastic approach (where no $G^\phi_0$ term is present), the sampling vectors are constructed as orthogonal to the $\{\phi\}$-subspace, i.e.:
\begin{equation}\label{projectedzeta}
    \left|\bar\zeta\right \rangle = \left( 1-P_\phi\right)  \left|\bar\zeta_0\right \rangle.
\end{equation} 
Here, $\bar\zeta_0$ spans in principle, the entire Hilbert space, and $P_\phi$ is:
\begin{equation}\label{projectorphi}
    P_\phi = \sum_{j\in \left\{\phi\right\} } \left| \phi_j\middle\rangle \middle \langle \phi_j \right|.
\end{equation}
The construction of $\tilde G_0$ remains the same as in the fully stochastic case:  $\tilde G_0$ is decomposed by a pair of vectors $\bar\zeta$ and $\zeta(t)$, cf.~Eqs.~\eqref{projectedzeta}, \eqref{tevolve} and \eqref{u_tprop}.

Note that it is possible to generalize the projector $P_\phi$, Eq.\eqref{projectorphi}, to an arbitrary $\left\{ \phi \right\}$-subspace.  The particular choice of $\phi$ does not affect the decomposition of the Green's function, Eq.~\eqref{G0decomposed},  or the time evolution of $\zeta(t)$. However, the dynamics of $G_0^\phi$ would require explicit action of $U_{0,t}$ on $\phi$ that is, in principle, not an eigenstate of $H_0$. We do not pursue this route here and select $\{\phi\}$-subspace composed from the starting point eigenstates.

\paragraph{Second,} the retarded induced potential $\tilde u$ is decomposed through partitioning of the time-dependent charge density, cf., Eq.~\eqref{eqn:u}, (omitting the space-time coordinates):
\begin{equation}\label{eqn:n_0}
n \equiv \tilde n + n_{\phi},
\end{equation}
where the $n_\phi$ is the density constructed from occupied states $\phi$ (which we assume to be mutually orthogonal):
\begin{equation}
    n_\phi(\br,t) = \sum_{j\in\{\phi\}} f_j \left| \phi_{j}(\br,t)\right|^2.
\end{equation}
Here, $f_j$ is the occupation of the $j^{\rm th}$ state. Note that choosing $\phi$ within the unoccupied subspace is meaningless in this context. The complementary part, $\tilde n$, is given by stochastic sampling,  Eq.~\eqref{TDdensity}, that employs random  vectors $\tilde \eta$:
\begin{equation}
    \left|\widetilde{\eta}\right \rangle = \left( 1-\tilde P_\phi\right) \left|\eta\right \rangle,
\end{equation}
where $\eta$ spans the entire occupied subspace and
\begin{equation}\label{projectorphi_occ}
    \tilde P_\phi = \sum_{j\in \left\{\phi\right\} } f_j \left| \phi_j\middle\rangle \middle \langle \phi_j \right|.
\end{equation}

The time propagation of the charge density is similar to the fully stochastic case. The deterministic and stochastic vectors evolve as:
\begin{align}
\left| \phi_j(t)\right\rangle &= U_{0,t} [n_\phi(t), \tilde n(t)] \left| \phi_j\right\rangle\\
    \left| \tilde \eta(t)\right\rangle &= U_{0,t} [n_\phi(t), \tilde n(t)] \left| \tilde \eta\right\rangle,
\end{align}
where $U_{0,t}$ is explicitly expressed as a functional of the two density contributions from  Eq.~\eqref{eqn:n_0}. Note that these expressions are analogous to Eq.~\eqref{tpropeta_Un}.

\paragraph{Third,} both partitionings are used in conjunction. While $G_0$ contains contributions from both occupied and unoccupied states, only the former are included in $\tilde u$. The combined partitioning may use entirely different subspaces for the Green's function and the induced potential. In Section~\ref{section:hBN}, we employ the decomposition in both terms because it yields the best results and significantly reduces statistical fluctuations.

\section{Decomposition of the stochastic polarization self-energy}
\label{sec:decomp_theory}

In the preceding section, we partition retarded induced potential and the Green's function, intending to decrease stochastic fluctuations in the self-energy. Here, we use the partitioning to achieve the second goal of this paper -- to determine the contribution to $\Sigma_P(\omega)$. 

Conceptually, we want to address quasiparticle scattering by correlations from a particular subspace. In the expression for $\Sigma_P(\omega)$, Eq.~\eqref{self-energy}, this corresponds to accounting for selected charge density fluctuations in $\tilde u$. In practice, we construct the \emph{subspace} polarization self-energy as:
\begin{equation}\label{sigma_subspace}
\langle\phi|\Sigma^s_P(t)|\phi\rangle = \frac{1}{N_{ \bar{\zeta}}} \sum_{\bar{\zeta}} \int \phi(\br)\zeta(\br,t) u^s_\zeta(\br,t) d^3\br,
\end{equation}
where we introduced the \emph{subspace} induced potential $u^s_\zeta$ which is obtained from its retarded form (in analogy to Eq.~\eqref{eqn:u}):
\begin{equation}
    \tilde u^s_\zeta(\br,t) =\int \nu(\br,\bp) \delta n^s(\bp,t) {\dd}\bp.
\end{equation}
This potential stems from the induced charge density that includes contributions only from selected orbitals $\{\phi\}$. Note that $n^s(\bp,t)$ is obtained either from individual single-particle states or from the stochastic sampling of the $\{\phi\}$-subspace under consideration. 

If the set of $\phi$ states is large, it is natural to employ the stochastic approach; the density is sampled  according to Eq.~\eqref{TDdensity} with vectors $\eta^s$ prepared as:
\begin{equation}
    \left| \eta^s \right \rangle = \tilde P_\phi \left| \eta \right \rangle
\end{equation}
where the projector is in Eq.~\eqref{projectorphi_occ} and vectors $\eta$ span the entire occupied subspace.

The time evolution of $\eta^s$ follows Eq.~\eqref{tpropeta_Un}, i.e., it is governed by the operator $U_{0,t}$ that depends on the \emph{total} time-dependent density:
\begin{equation}\label{eta_s_tprop}
    \left|  \eta^s(t)\right\rangle = U_{0,t} [n(t)] \left|  \eta^s\right\rangle.
\end{equation}
Hence, despite $\Sigma_P^s(t)$ contains only fluctuations from a particular subspace, the calculation requires knowledge of the time evolution of both $n^s$ and $n$.

In practice, we employ a set of two independent stochastic samplings: (i) vectors $\left|\eta\right\rangle$ describing the entire occupied space, and (ii) vectors  $\left|\eta^s\right\rangle$ confined only to the chosen $\{\phi\}$-subspace. The first set characterizes the total change density fluctuation and enters $U_{0,t}$ in Eq.~\ref{eta_s_tprop}.

\section{Numerical results and discussion}
\subsection{Computational details}

In this section, we will demonstrate the capabilities of the method introduced above.
The starting-point calculations are performed with a real-space DFT implementation, employing regular grids, Troullier-Martins pseudopotentials,\cite{TroullierMartins1991} and the PBE\cite{PerdewWang} functional for exchange and correlation. We investigate finite and 2D infinite systems using modified periodic boundary conditions with Coulomb interaction cutoffs.\cite{Rozzi_2006}

The numerical verification for the SiH$_4$ molecule is in section \ref{sec:sih4}. To converge the occupied $H_0$ eigenvalues to $< 5$~meV, we use a kinetic energy cutoff of 26~$E_h$ and $64\times 64 \times 64$ real-space grid with the step of 0.3~$a_0$.

Large calculations for the $V_N$ defect in hBN monolayer and in hBN heterostructure with graphene are in sections \ref{section:hBN} and \ref{sec:vn_heterostructure}. In both cases, we consider relaxed rectangular 12$\times$6 supercells containing 287 and 575 atoms. We performed a structural optimization in QuantumEspresso code\cite{QE2017} together with Tkatchenko-Scheffler's total energy corrections.\cite{TS_2009} The heterostructure is built with an interlayer distance of 3.35~\AA. 

The $GW$ calculations were performed using a development version of the StochasticGW code.\cite{neuhauser2014breaking, Vlcek2018swift, vlcek2017stochastic}
The calculations employ an additional set of 20,000 random vectors used in the sparse stochastic compression used for time-ordering of $\tilde u$\cite{Vlcek2018swift}.
The time propagation of the induced charge density is performed for maximum propagation time of 50~a.u., with the time-step of 0.05~a.u.

\subsection{Verification using molecular states}
\label{sec:sih4}
\begin{figure}
  \includegraphics[width=3.37in]{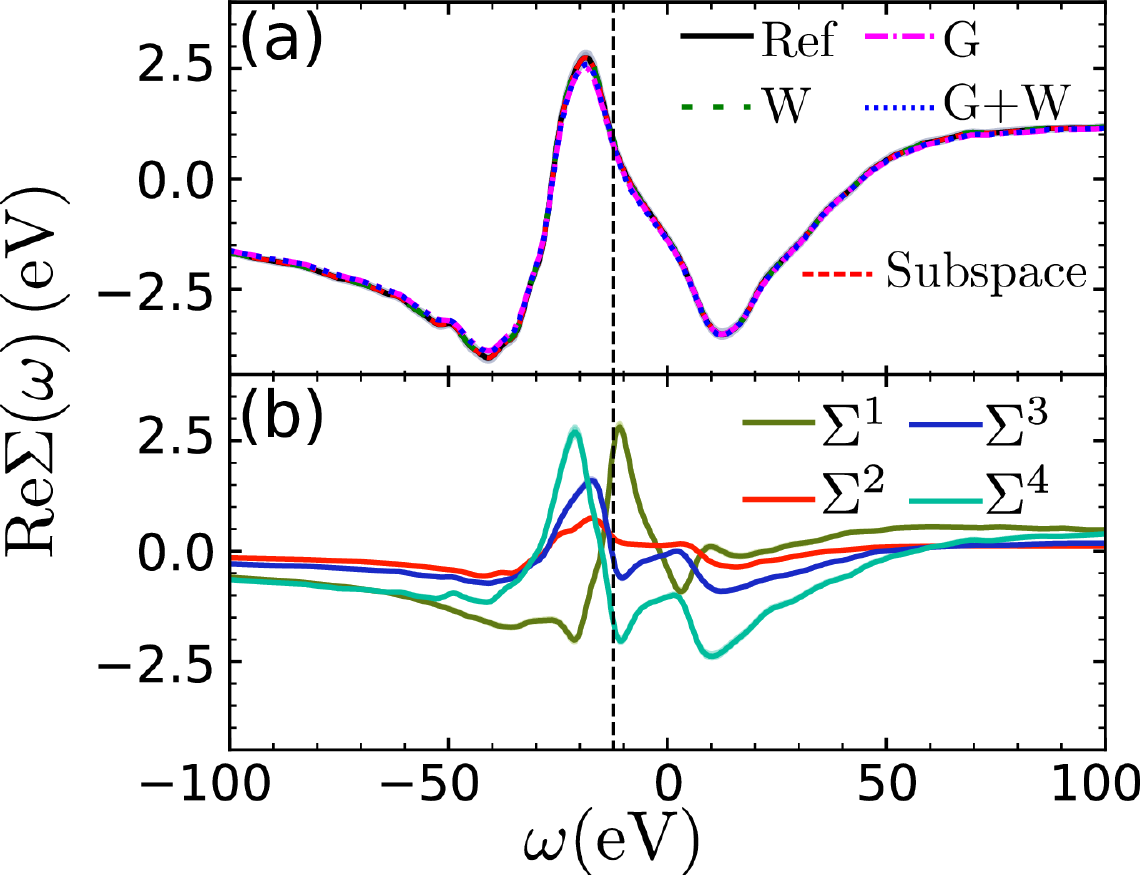}
  \caption{Verification for the SiH$_4$ molecule. (a)~The comparison of the total self-energies of the highest occupied state (HOMO) obtained with reference and embedding schemes. The figure demonstrates that all five approaches yield identical $\Sigma(\omega)$. In the plot, G denotes the decomposition scheme where HOMO is embedded in the Green's function; W denotes the calculation with HOMO state embedded in screened Coulomb potential; G+W corresponds to simultaneous decomposition of G and W. The red-dashed line represents the sum over subspace contributions shown in the panel below. (b)~Subspace self-energy of the HOMO state that contains only a contribution of a specific state $i$, denoted as $\Sigma^i$ in the figure's legend.}
  \label{fgr:SiH4}
\end{figure}

We verify the implementation of the methods presented in Sections~\ref{sec:embed_theory} and \ref{sec:decomp_theory} on SiH$_4$ molecule. In particular, we test separately : (i) the construction of the embedding schemes, (ii) decomposition of the time propagation, and (iii) the evaluation of the subspace self-energy. 

The reference calculation employs only one level of stochastic sampling (for the Green's function, while the rest of the calculation is deterministic). 
For small systems, this approach converges fast as the stochastic fluctuations are small.\cite{Vlcek2018swift, vlcek2017stochastic} Yet, we need $N_\zeta = 1,500$ vectors to decrease the QP energies errors below 0.08~eV. An illustration of the self-energy for the HOMO state of SiH$_4$ is in Fig.~\ref{fgr:SiH4}; everywhere in the figure, the stochastic error is below $0.13$~eV for all frequencies.

We first inspect the results for an embedded deterministic subspace.  Fig.~\ref{fgr:SiH4} shows that the different schemes for the explicit treatment of the HOMO state, Section~\ref{sec:embed_theory}, produce the same self-energy curve. 
The inclusion of HOMO in $W$ is combined with the stochastic sampling of the three remaining orbitals by $N_\eta=$16 random vectors.  Note that it is not economical to sample the action of $W_P$ for small systems,\cite{vlcek2017stochastic,Vlcek2018swift} and this calculation serves only as a test case. This treatment yields a statistical error of 0.08~eV, i.e., the same as the reference calculation despite the additional fluctuations due to the $\eta$ vectors. 

When the HOMO orbital is explicitly included in $G$, the resulting statistical error is below the error of the reference calculation (0.05~eV). Such a result is expected since the reference relies on the stochastic sampling of the Green's function. The same happens when the frontier orbital is in both $G$ and $W$ ($N_\eta=$16 random vectors sample the induced charge density). Tests for other states are not presented here but lead to identical conclusions.

Next, we verify that the density entering the time-evolution operator $U_{0,t} [n(t)]$ can be constructed from different states than it is acting on. Namely, the induced charge and the time-dependent densities may be sampled and built by different means. To demonstrate this, we propagate each of the $H_0$ eigenstates with $U_{0,t} [n(t)]$, where $n(t)$ is \emph{stochastically sampled} by $N_\eta$=16 random vectors. Only the induced charge density, entering Eq.~\eqref{eqn:u}, is computed from the $\{\phi\}$ eigenvectors. The agreement with the reference self-energy curve is excellent with differences smaller than the standard deviations at each frequency point; see  Fig.~\ref{fgr:SiH4}a.

Finally, we inspect the subspace self-energy in which  $U_{0,t} [n(t)]$ employs the total charge density sampled by $N_\eta$=16 random vectors. In Fig.~\ref{fgr:SiH4}b shows four different $\Sigma^s(\omega)$ curves corresponding to the contributions of individual $H_0$ eigenstates. Since the eigenstates are orthogonal, the total self-energy is simply the sum of individual $\Sigma^s(\omega)$ components. The additivity of the subspace self-energies is demonstrated numerically in Fig.~\ref{fgr:SiH4}a. The subspace results illustrate that HOMO and the bottom valence orbital exhibit the largest amplitudes of $\Sigma_P^s$; hence, these two states dominate the correlation near the ionization edge.

\subsection{Deterministic treatment of localized states}
\label{section:hBN}

The deterministic subspace embedding should numerically stabilize the stochastic sampling and decreases the computational cost. To test the methodology on a realistic system, we consider the electronic structure of an infinitely periodic hBN monolayer containing a single nitrogen vacancy ($V_N$) per a unit cell with dimensions of $3.0\times 2.6$~nm. The system comprises 1147 electrons with the defect state being singly occupied and hence positioned at the Fermi level. 

In the current calculations, we enforce spin degeneracy. The reason is twofold: (i) half populated states are strongly polarizable, and they exhibit stronger stochastic fluctuations in the time evolution. They are thus a more stringent test of the embedding. (ii) In Section~\ref{sec:vn_heterostructure}, we compare the monolayer with a heterostructure to determine the role of substrate material on the self-energy. In the heterostructure, the magnetic splitting of spin-up/down components disappears.\cite{Park_2014}

\begin{figure} 
  \includegraphics[width=3.37in]{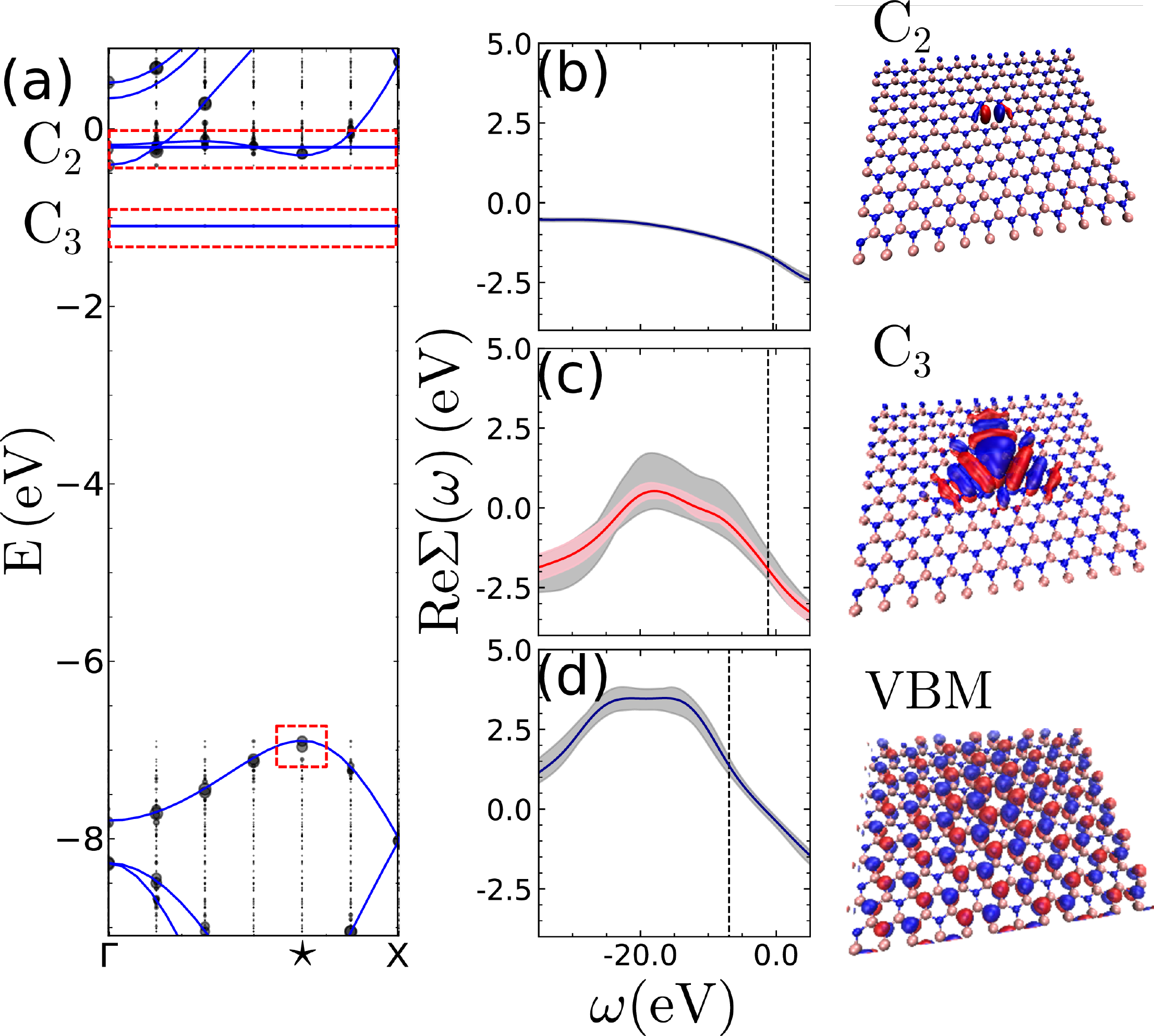}
  \caption{Deterministic embedding for the nitrogen vacancy in hBN monolayer. (a) Black points represent individual states the unfolded band structure; blue lines are a guide for the eyes. Panels b, c, and d depict the self-energy curves (solid line) for $C_2$, $C_3$, and VBM state respectively; the states are marked in panel a by red dashed rectangles. The statistical errors are smaller than the thickness of the line. Grey shaded areas (on each plot) correspond to the ``spread'' of self-energy curves for 15 distinct fully stochastic calculations; solid blue lines are the statistical averages. The light red area in panel c represents the spread after deterministic embedding of the $C_3$ state; the solid red line is the self-energy curve.  The third column contains the electron density of the corresponding states.}
  \label{fgr:hBN} 
\end{figure}

The relaxed monolayer geometry shows only mild restructuring. The vacancy introduces three localized states with $C_3$ and $C_2$ spatial symmetry.  The former ($C_3$) is singly occupied and forms an in-gap state. The latter state is doubly degenerate and pushed high in the conduction region. Due to the enforced spin-degeneracy, the electron-electron interactions are increased and $C_2$ appears higher than in the previous calculations\cite{Attaccalite_2011,Huang_2012, Tran_2016,McDougall_2017}.  The $C_3$ and $C_2$ single-particle wavefunctions are illustrated in Fig.~\ref{fgr:hBN}.

We first focus on the delocalized top valence and bottom conduction states. The fully stochastic calculations converge fast for both of them. The charge density fluctuations are sampled by $N_\eta=8$ vectors, and the Green's function requires $N_{{\zeta}}$=1500 to yield QP energies with statistical errors below 0.03~eV for the valence band maximum (VBM). The error for the conduction band minimum (CBM) is $<0.01$~eV. The computed resulting quasiparticle band-gap is $6.49\pm0.04$~eV, in excellent agreement with previous calculations and experiments (providing a range of values between 6.1 and 6.6~eV).\cite{Levinshtein_2001,Fuchs_2007,Huser_2013}

To investigate the electronic structure in greater detail, we employ the projector-based energy-momentum analysis based on supercell band unfolding.\cite{Popescu_2012, Huang_2014, Medeiros_2014,Brooks_2020} In practice, the individual wavefunctions within our simulation cell are projected onto the Brillouin zone of a single hBN unit cell. The resulting bandstructure is shown in Fig.~\ref{fgr:hBN}a. Since our calculations employ a \emph{rectangular} supercells, the critical point K of the hexagonal Brillouin zone appears between the $\Gamma$ and X points (marked on the horizontal axis by $\star$). The figure shows that the fundamental band gap is indirect; the direct transition is $7.39\pm0.04$ eV in extremely close to the results for the pure hBN monolayer (reported to be in a range between 7.26 and 7.37~eV ).\cite{Huser_2013,Cudazzo_2016,Paleari_2018}

The defect states appear as flat bands, labeled in Fig.~\ref{fgr:hBN} by their symmetry. As expected from the outset, the stochastic calculations for $C_3$ exhibit large fluctuations. While each sample is numerically stable, random vectors produce a self-energy curve with a significant statistical error at each frequency point, i.e., the time evolution is strongly dependent on the initial choice of $\{\zeta\}$ and $\{\eta\}$. In this example, only the $C_3$ state exhibits such behavior.

In Fig.~\ref{fgr:hBN} illustrates the self-energies for the band edge and the defect states. For all the cases, the plots show the \emph{spread} of the $\Sigma_P(\omega)$ curves: these correspond to the outer envelope for 15 distinct calculations, each employing 100 $\zeta$ sampling vectors (combined with $N_\eta =8$ each). For the $C_3$ defect state, the stochastic sampling is possibly biased. The variation is three times as big as the spread of the VBM, and almost seven times larger compared to CBM.   Away from the QP energy, the fluctuations increase even further; the spread of the samples becomes two times larger near the maximum of $\Sigma_P$ at $-20$~eV.  The convergence of the QP energy is poor, and the low sample standard deviation (roughly twice as big as for VBM) suggests incorrect statistics. In practice, each sampling yields a self-energy curve that lies outside of the standard deviation of the previous simulation.

The deterministic embedding remedies insufficient sampling without increasing the computational cost. Naturally, we select the $C_3$ defect state and treat it explicitly (while randomly sampling the rest of the orbitals). Hence, according to the notation of Section~\ref{sec:embed_theory}, the $\{\phi\}$-subspace contains only a single orbital.

The decomposition of the non-interacting Green's function follows Eq.~\eqref{G0decomposed} and stabilizes the sampling. The spread of the self-energy curves decreases approximately three times for a wide range of frequencies. With the embedded $C_3$ state, the statistical error of the QP energy decreases smoothly and uniformly with the number of samples. Each new sampling falls within the error of the calculations. Yet,  the final statistical error (0.03~eV) is less than 10\% larger than for the delocalized states.

The decomposition of the induced charge density alone is less promising. Fundamentally, $\delta n(\br,t)$ contains contributions from the entire system, and a small $\{\phi\}$-subspace will unlikely lead to drastic improvement. Indeed, the statistical errors and convergence behavior remain the same as for the fully stochastic treatment. 

Embedding of the localized state in both $G_0$ and $\delta n(\br,t)$ is, however, the best strategy. If both decompositions use the same (or overlapping) $\{\phi\}$-subspace, the induced charge density and the potential $\delta v$, Eq.~\eqref{perturbing_pot}, share (at least some) $\phi$ states. Consequently, the sampling of $\Sigma_P(\omega)$ becomes less dependent on the particular choice of $\zeta$ vectors, which sample only states orthogonal to $\{\phi\}$. Indeed, the embedding of a single localized state in $G$ and $W$ results in a nearly \emph{four-fold reduction of the statistical fluctuation} that translates to more than an order of magnitude savings in the computational time. The error of the QP energy of the $C_3$ state is approximately half of the error for VBM (16~meV for $N_\zeta = 1,500$).  

For completeness, we also applied the three types of embedding on VBM, which does not suffer from bias or large statistical errors. Unlike for the localized state, we observe only a negligible reduction of the fluctuations. For delocalized states, the fully stochastic sampling is thus sufficient as expected from our previous work.\cite{Brooks_2020,Vlcek2018swift, vlcek2017stochastic}

\subsection{Stochastic subspace self-energy}\label{sec:vn_heterostructure}

Having an improved description of the localized states in hand, we now turn to the decomposition of the self-energy. The real-time approach described in Section~\ref{sec:decomp_theory} allows inspecting the many-body interactions from a selected portion of the system. In contrast to the previous subsection, the subspace of interest contains a large number of states (irrespective of their degree of localization), and it is randomly sampled. The goal of this decomposition is to understand the role of correlation, especially at interfaces.

\begin{figure*} 
\includegraphics[width=6.69in]{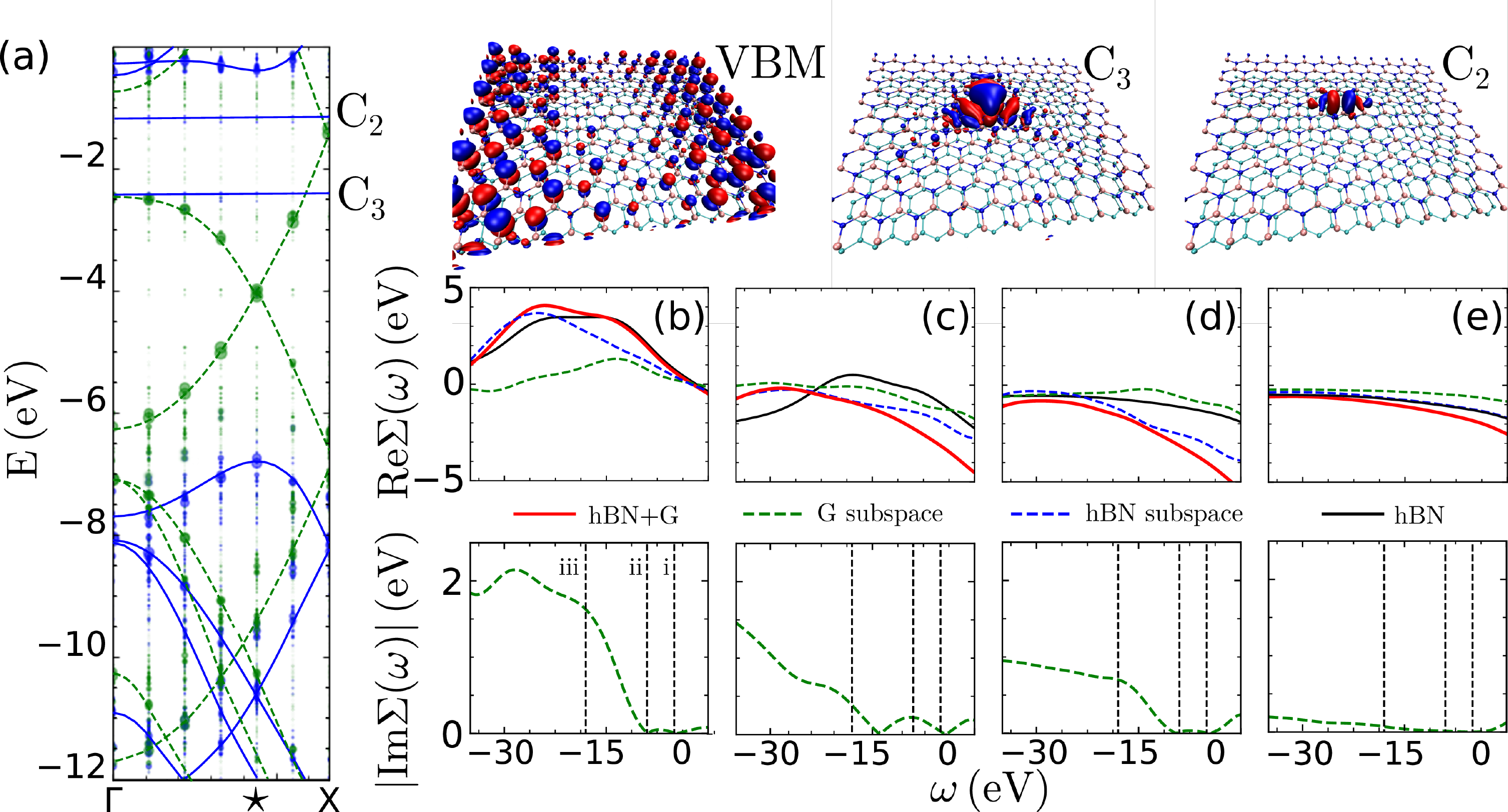}
\caption{Excitations in an hBN-graphene heterostructure containing a nitrogen vacancy. (a) Calculated band structure -- the states localized on graphene are in green, while blue points represent hBN. Lines serve as a guide for the eyes. The individual isosurfaces are shown on the right for VBM, $C_3$ defect state, and $C_2$ defect states. Second row contains averaged self-energy curves of the: (b) VBM, (c) $C_3$ defect state, (d) $C_2$ defect state, (e) CBM. Solid red lines are the full heterostructure self-energies; green and blue dashed lines are respectively the $\Sigma^s_P(\omega)$ graphene monolayer and hBN monolayer subspace self-energies. The black line represents the self-energy of an isolated hBN monolayer with $V_N$. The third row of panels contains the $\mathrm{Im}\Sigma^s_P(\omega)$ of graphene. Dashed vertical lines indicate the features in the experimental EELS spectrum from  Ref.~\onlinecite{Kinyanjui_2012}. The experimentally observed features, labeled as i, ii, and iii, are attributed to $\pi \rightarrow \pi*$ interband transitions, $\pi$, and $\pi + \sigma$ plasmon excitations, respectively.}
\label{fgr:hBN_gphn} 
\end{figure*}

To test our method, we investigate a periodic hBN monolayer containing a single $V_N$ defect placed on graphene. Such heterostructure has also been realized experimentally. 
\cite{Xu_2020, Salihoglu_2016, Lee_2014, Bjelkevig_2010} 
The structure contains 2299 valence electrons, and it is illustrated in Fig.~\ref{fgr:hBN_gphn} together with selected orbital isosurfaces. The $C_3$ state is energetically lower than $C_2$ (as in the pristine hBN monolayer). Both defects only weakly hybridize with graphene and remain localized within the hBN sublayer. However, graphene presence leads to a slightly increased delocalization of the defects within the monolayer.\footnote{The distributions of the wavefunctions are governed by the local external and Hartree potentials (since the orbitals correspond to eigenstates of the mean-field Hamiltonian $H_0$).}

By unfolding the wave functions of the bilayer, we obtain the band structure shown in Fig.~\ref{fgr:hBN_gphn}a. The graphene and hBN bands are distinguished by a  state projection on the densities above and below the center of the interlayer region. Note that this simple approach captures hybridized states as having dual character (i.e., they appear having both graphene and hBN contributions). 

The graphene portion of the band structure reproduces the well-known semimetallic features with a Dirac point located at the K boundary of the hexagonal Brillouin zone. As discussed above, the K appears in between $\Gamma$ and X of the rectangular cell, and it is labeled by $\star$ in Fig.~\ref{fgr:hBN_gphn}. The typical Dirac cone dispersion of graphene is only a little affected by the hBN presence. However, the ordering of the hBN and graphene states is nontrivial. In contrast to previous DFT calculations for 3-times higher defect density, we notice that the Dirac point remains close to the Fermi level despite the charge transfer from the $C_3$ defect state.\cite{Park_2014} This is not surprising: sparse charge defects lead to only weak doping of graphene. Further, the previous DFT results employed small supercells and may suffer from significant electronic ``overdelocalization''\cite{Mori_Sanchez_2008,Cohen_2008} that spuriously enhances charge transfer.

The hBN part of the band structure,  Fig.~\ref{fgr:hBN_gphn}, is only weakly affected by the heterostructure formation. The delocalized states are qualitatively identical to those in the monolayer, Fig.~\ref{fgr:hBN}. The fundamental band-gap remains indirect and reduced to $6.32\pm0.04$~eV. The screening introduced by graphene thus leads to a small change in $E_g$ ($<0.2$~eV compared to the monolayer). The positions of the defect states, however, change notably. Here, both appear above the Fermi level.

Due to the charge transfer and Fermi level shift, it is clear that graphene is responsible for altering the charge fluctuations, i.e., the polarization part of the self-energy. In practice, graphene acts as a dielectric background inducing a significant screening of the Coulomb interaction in hBN. It stands to reason that the localized defect states would be strongly affected by such a polarizable layer and that the corresponding self-energy should be dominated by the spectral features originating in graphene. 

To investigate the degree of coupling and the contribution to the self-energy from each monolayer, we compute $\Sigma^s_P(\omega)$ using Eq.~\eqref{sigma_subspace}. Here,  the $\{\phi\}$-subspace is constructed from stochastic samples of the  576 occupied graphene states (distinguished by green color in  Fig.~\ref{fgr:hBN_gphn}). For each sampling of the Green's function, the subspace is described by eight random vectors in the $\{\phi\}$-subspace. Additional eight vectors sample the complementary subspace.

Fig.~\ref{fgr:hBN_gphn} shows the decomposition of the self-energy, indicating the contribution of graphene. For the delocalized states, the $\Sigma_P(\omega)$ curves appear similar to those in the hBN monolayer. Specifically, we see that $\Sigma_P(\omega)$ for VBM  has practically identical frequency dependence between -15~and~0~eV. While in-plane contributions from hBN dominate the entire curve, screening from the graphene substrate is significant. A na\"ive comparison between the monolayer hBN and the heterostructure suggests that the enhanced maximum in Re$\Sigma_P$ at $-25$~eV is caused extrinsically by graphene. Surprisingly, this is not the case, and the effect is only indirect: the presence of graphene leads to shifting of the hBN spectral features. At the QP energy, graphene contributes to the correlation by  $36\%$.

The situation is different for CBM. Towards the static limit ($\omega\to0$), the self-energy curve becomes more negative. This shift is caused directly by the induced density fluctuations in the graphene substrate. The decreased polarization self-energy indicates the stabilization of the conduction states, which is directly linked to the attractive van der Waals fluctuations in bilayer systems.\cite{Brooks_2020}

The self-energy curves of the $C_3$ and $C_2$ defect states are significantly different from those computed for the monolayer. In both cases, we observe a negative shift related to the QP stabilization by non-local correlations. The $\Sigma_P$ contributions to the defects' QP energies are roughly three times larger in the heterostructure.
Na\"ively, we expect that the localized states are strongly coupled to graphene polarization modes, which lead to negative $\Sigma_P$. Surprisingly, the intrinsic hBN interactions are the major driving force. The defect states are mostly affected by the in-plane induced density fluctuations (constituting  $64\%$ and $76\%$ of the total polarization self-energy for  $C_3$ and  $C_2$). Like the VBM, graphene indirectly acts on the defect QP states by enhancing the in-plane fluctuations, rather than direct coupling to the defects.

For all the states illustrated in Fig.~\ref{fgr:hBN_gphn}, the graphene subspace contributions are governed by density oscillations induced by electron removal from or addition to the hBN layer. Collective charge excitations, i.e., plasmons, naturally correspond to the most prominent features. They are the poles of the screened Coulomb interaction and appear as strong peaks in the imaginary part of the subspace self-energy (bottom panels in Fig.~\ref{fgr:hBN_gphn}).

Conceptually, the substrate plasmons are directly related to QP energy dissipation. Due to the weak coupling between the monolayers, Im$\Sigma_P^s(\omega)$ of the substrate shows the same spectral features at the experimental electron energy loss spectra measured for graphene alone. As distinct states couple to individual plasmon modes differently, the intensity of peaks in  Im$\Sigma_P^s(\omega)$ varies. Due to changes in geometry and the presence of hBN, the spectral features are shifted from the experimental data. Still, we identify $\pi+\sigma$ and $\pi$ plasmons. The former is universally the most prominent feature in the subspace self-energy. The $\pi$ (possibly together with $\pi\to \pi^*$) plasmon contributes much less, and it is appreciable only for the $C_3$ defect state. We surmise that the additional peak near zero frequency corresponds to the ``static'' polarization of graphene due to the charge transfer from the defect. 

As shown in this example, the stochastic subspace self-energy efficiently probes dynamical electron-electron interactions at interfaces. The random sampling allows selecting an arbitrary subspace to identify its contribution to the correlation energy and the excited state lifetimes.

\section{Conclusions}

Stochastic Green'€™s function approaches represent a class of low-scaling many-body methods based on stochastic decomposition and random sampling of the Hilbert space. The overall computational cost is impacted by the presence of localized states that increase the sampling errors.  
Here, we have introduced a practical solution for the $G_0W_0$ method. The localized states are treated explicitly (deterministically), while the rest is subject to stochastic sampling. We have further shown that the subspace-separation can be applied to decompose the dynamical self-energy into contributions from distinct states (or parts of the systems). 

Using nitrogen vacancy in hBN monolayer, we demonstrate that the deterministic embedding dramatically reduces the statistical fluctuations. Consequently,  the computational time decreases by more than an order of magnitude for a given target level of stochastic error. The embedding is made within the Green's function and the screened Coulomb interactions independently or simultaneously; the latter provides the best results. For delocalized states, embedding is not necessary and performs similar to the standard uniform sampling with real-space random vectors. The embedding scheme presented here is general, and it is the first step towards hybrid techniques that treat selected subspace at the distinct (higher) level of theory. The development of hybrid techniques is currently underway.

We further demonstrate that the subspace self-energy contains (in principle additive) contributions from the induced charge density. The self-energy contribution from a particular portion of the system is computed by confining sampling vectors into different (orthogonal) subspaces. We exemplify the capabilities of such calculations on defects state in the hBN-graphene heterostructure. Here, the charged excitation in one layer directly couples to density oscillations in the substrate monolayer. The response is dominated by plasmon modes, which are identified as features in the imaginary part of the subspace self-energy. Surprisingly, the electronic correlations for the defects are governed by the interactions in the host layer; the substrate only indirectly affects their strength. 

This example serves as a stimulus for additional study of the defect states in heterostructures. Here, the coupling strength between individual subsystems is tunable by particular stacking order and induced strains. The subspace self-energy represents a direct route to explore such quantum many-body interfacial phenomena.

\begin{acknowledgements}

This work was supported by the NSF through the Materials
Research Science and Engineering Centers (MRSEC)
Program of the NSF through Grant No. DMR-1720256 (Seed
Program).  In part, this work was supported by the NSF Quantum Foundry through Q-AMASE-i
program Award No. DMR-1906325. The calculations were performed as part of the XSEDE\cite{Towns_2014} computational Project No.~TG-CHE180051. Use was made of computational facilities purchased with
funds from the National Science Foundation (CNS-1725797)
and administered by the Center for Scientific Computing
(CSC). The CSC is supported by the California NanoSystems
Institute and the Materials Research Science and Engineering
Center (MRSEC; NSF DMR-1720256) at UC Santa Barbara. 

\end{acknowledgements}

\bibliography{biblio}

\begin{thebibliography}{61}%
\makeatletter
\providecommand \@ifxundefined [1]{%
 \@ifx{#1\undefined}
}%
\providecommand \@ifnum [1]{%
 \ifnum #1\expandafter \@firstoftwo
 \else \expandafter \@secondoftwo
 \fi
}%
\providecommand \@ifx [1]{%
 \ifx #1\expandafter \@firstoftwo
 \else \expandafter \@secondoftwo
 \fi
}%
\providecommand \natexlab [1]{#1}%
\providecommand \enquote  [1]{``#1''}%
\providecommand \bibnamefont  [1]{#1}%
\providecommand \bibfnamefont [1]{#1}%
\providecommand \citenamefont [1]{#1}%
\providecommand \href@noop [0]{\@secondoftwo}%
\providecommand \href [0]{\begingroup \@sanitize@url \@href}%
\providecommand \@href[1]{\@@startlink{#1}\@@href}%
\providecommand \@@href[1]{\endgroup#1\@@endlink}%
\providecommand \@sanitize@url [0]{\catcode `\\12\catcode `\$12\catcode
  `\&12\catcode `\#12\catcode `\^12\catcode `\_12\catcode `\%12\relax}%
\providecommand \@@startlink[1]{}%
\providecommand \@@endlink[0]{}%
\providecommand \url  [0]{\begingroup\@sanitize@url \@url }%
\providecommand \@url [1]{\endgroup\@href {#1}{\urlprefix }}%
\providecommand \urlprefix  [0]{URL }%
\providecommand \Eprint [0]{\href }%
\providecommand \doibase [0]{http://dx.doi.org/}%
\providecommand \selectlanguage [0]{\@gobble}%
\providecommand \bibinfo  [0]{\@secondoftwo}%
\providecommand \bibfield  [0]{\@secondoftwo}%
\providecommand \translation [1]{[#1]}%
\providecommand \BibitemOpen [0]{}%
\providecommand \bibitemStop [0]{}%
\providecommand \bibitemNoStop [0]{.\EOS\space}%
\providecommand \EOS [0]{\spacefactor3000\relax}%
\providecommand \BibitemShut  [1]{\csname bibitem#1\endcsname}%
\let\auto@bib@innerbib\@empty
\bibitem [{\citenamefont {Yu}\ \emph {et~al.}(2017)\citenamefont {Yu},
  \citenamefont {Liu}, \citenamefont {Tang}, \citenamefont {Xu},\ and\
  \citenamefont {Yao}}]{Yu_2017}%
  \BibitemOpen
  \bibfield  {author} {\bibinfo {author} {\bibfnamefont {H.}~\bibnamefont
  {Yu}}, \bibinfo {author} {\bibfnamefont {G.-B.}\ \bibnamefont {Liu}},
  \bibinfo {author} {\bibfnamefont {J.}~\bibnamefont {Tang}}, \bibinfo {author}
  {\bibfnamefont {X.}~\bibnamefont {Xu}}, \ and\ \bibinfo {author}
  {\bibfnamefont {W.}~\bibnamefont {Yao}},\ }\bibfield  {title} {\enquote
  {\bibinfo {title} {Moir{\'e} excitons: From programmable quantum emitter
  arrays to spin-orbit{\textendash}coupled artificial lattices},}\ }\href
  {\doibase 10.1126/sciadv.1701696} {\bibfield  {journal} {\bibinfo  {journal}
  {Science Advances}\ }\textbf {\bibinfo {volume} {3}} (\bibinfo {year}
  {2017}),\ 10.1126/sciadv.1701696}\BibitemShut {NoStop}%
\bibitem [{\citenamefont {Grosso}\ \emph {et~al.}(2017)\citenamefont {Grosso},
  \citenamefont {Moon}, \citenamefont {Lienhard}, \citenamefont {Ali},
  \citenamefont {Efetov}, \citenamefont {Furchi}, \citenamefont
  {Jarillo-Herrero}, \citenamefont {Ford}, \citenamefont {Aharonovich},\ and\
  \citenamefont {Englund}}]{Grosso2017}%
  \BibitemOpen
  \bibfield  {author} {\bibinfo {author} {\bibfnamefont {G.}~\bibnamefont
  {Grosso}}, \bibinfo {author} {\bibfnamefont {H.}~\bibnamefont {Moon}},
  \bibinfo {author} {\bibfnamefont {B.}~\bibnamefont {Lienhard}}, \bibinfo
  {author} {\bibfnamefont {S.}~\bibnamefont {Ali}}, \bibinfo {author}
  {\bibfnamefont {D.~K.}\ \bibnamefont {Efetov}}, \bibinfo {author}
  {\bibfnamefont {M.~M.}\ \bibnamefont {Furchi}}, \bibinfo {author}
  {\bibfnamefont {P.}~\bibnamefont {Jarillo-Herrero}}, \bibinfo {author}
  {\bibfnamefont {M.~J.}\ \bibnamefont {Ford}}, \bibinfo {author}
  {\bibfnamefont {I.}~\bibnamefont {Aharonovich}}, \ and\ \bibinfo {author}
  {\bibfnamefont {D.}~\bibnamefont {Englund}},\ }\bibfield  {title} {\enquote
  {\bibinfo {title} {Tunable and high-purity room temperature single-photon
  emission from atomic defects in hexagonal boron nitride},}\ }\href {\doibase
  10.1038/s41467-017-00810-2} {\bibfield  {journal} {\bibinfo  {journal}
  {Nature Communications}\ }\textbf {\bibinfo {volume} {8}},\ \bibinfo {pages}
  {705} (\bibinfo {year} {2017})}\BibitemShut {NoStop}%
\bibitem [{\citenamefont {Yankowitz}\ \emph {et~al.}(2018)\citenamefont
  {Yankowitz}, \citenamefont {Jung}, \citenamefont {Laksono}, \citenamefont
  {Leconte}, \citenamefont {Chittari}, \citenamefont {Watanabe}, \citenamefont
  {Taniguchi}, \citenamefont {Adam}, \citenamefont {Graf},\ and\ \citenamefont
  {Dean}}]{Yankowitz2018}%
  \BibitemOpen
  \bibfield  {author} {\bibinfo {author} {\bibfnamefont {M.}~\bibnamefont
  {Yankowitz}}, \bibinfo {author} {\bibfnamefont {J.}~\bibnamefont {Jung}},
  \bibinfo {author} {\bibfnamefont {E.}~\bibnamefont {Laksono}}, \bibinfo
  {author} {\bibfnamefont {N.}~\bibnamefont {Leconte}}, \bibinfo {author}
  {\bibfnamefont {B.~L.}\ \bibnamefont {Chittari}}, \bibinfo {author}
  {\bibfnamefont {K.}~\bibnamefont {Watanabe}}, \bibinfo {author}
  {\bibfnamefont {T.}~\bibnamefont {Taniguchi}}, \bibinfo {author}
  {\bibfnamefont {S.}~\bibnamefont {Adam}}, \bibinfo {author} {\bibfnamefont
  {D.}~\bibnamefont {Graf}}, \ and\ \bibinfo {author} {\bibfnamefont {C.~R.}\
  \bibnamefont {Dean}},\ }\bibfield  {title} {\enquote {\bibinfo {title}
  {Dynamic band-structure tuning of graphene moir{\'e} superlattices with
  pressure},}\ }\href {\doibase 10.1038/s41586-018-0107-1} {\bibfield
  {journal} {\bibinfo  {journal} {Nature}\ }\textbf {\bibinfo {volume} {557}},\
  \bibinfo {pages} {404--408} (\bibinfo {year} {2018})}\BibitemShut {NoStop}%
\bibitem [{\citenamefont {Cao}\ \emph {et~al.}(2018)\citenamefont {Cao},
  \citenamefont {Fatemi}, \citenamefont {Demir}, \citenamefont {Fang},
  \citenamefont {Tomarken}, \citenamefont {Luo}, \citenamefont
  {Sanchez-Yamagishi}, \citenamefont {Watanabe}, \citenamefont {Taniguchi},
  \citenamefont {Kaxiras}, \citenamefont {Ashoori},\ and\ \citenamefont
  {Jarillo-Herrero}}]{Cao2018}%
  \BibitemOpen
  \bibfield  {author} {\bibinfo {author} {\bibfnamefont {Y.}~\bibnamefont
  {Cao}}, \bibinfo {author} {\bibfnamefont {V.}~\bibnamefont {Fatemi}},
  \bibinfo {author} {\bibfnamefont {A.}~\bibnamefont {Demir}}, \bibinfo
  {author} {\bibfnamefont {S.}~\bibnamefont {Fang}}, \bibinfo {author}
  {\bibfnamefont {S.~L.}\ \bibnamefont {Tomarken}}, \bibinfo {author}
  {\bibfnamefont {J.~Y.}\ \bibnamefont {Luo}}, \bibinfo {author} {\bibfnamefont
  {J.~D.}\ \bibnamefont {Sanchez-Yamagishi}}, \bibinfo {author} {\bibfnamefont
  {K.}~\bibnamefont {Watanabe}}, \bibinfo {author} {\bibfnamefont
  {T.}~\bibnamefont {Taniguchi}}, \bibinfo {author} {\bibfnamefont
  {E.}~\bibnamefont {Kaxiras}}, \bibinfo {author} {\bibfnamefont {R.~C.}\
  \bibnamefont {Ashoori}}, \ and\ \bibinfo {author} {\bibfnamefont
  {P.}~\bibnamefont {Jarillo-Herrero}},\ }\bibfield  {title} {\enquote
  {\bibinfo {title} {Correlated insulator behaviour at half-filling in
  magic-angle graphene superlattices},}\ }\href {\doibase 10.1038/nature26154}
  {\bibfield  {journal} {\bibinfo  {journal} {Nature}\ }\textbf {\bibinfo
  {volume} {556}},\ \bibinfo {pages} {80--84} (\bibinfo {year}
  {2018})}\BibitemShut {NoStop}%
\bibitem [{\citenamefont {Zondiner}\ \emph {et~al.}(2020)\citenamefont
  {Zondiner}, \citenamefont {Rozen}, \citenamefont {Rodan-Legrain},
  \citenamefont {Cao}, \citenamefont {Queiroz}, \citenamefont {Taniguchi},
  \citenamefont {Watanabe}, \citenamefont {Oreg}, \citenamefont {von Oppen},
  \citenamefont {Stern}, \citenamefont {Berg}, \citenamefont
  {Jarillo-Herrero},\ and\ \citenamefont {Ilani}}]{Zondiner2020}%
  \BibitemOpen
  \bibfield  {author} {\bibinfo {author} {\bibfnamefont {U.}~\bibnamefont
  {Zondiner}}, \bibinfo {author} {\bibfnamefont {A.}~\bibnamefont {Rozen}},
  \bibinfo {author} {\bibfnamefont {D.}~\bibnamefont {Rodan-Legrain}}, \bibinfo
  {author} {\bibfnamefont {Y.}~\bibnamefont {Cao}}, \bibinfo {author}
  {\bibfnamefont {R.}~\bibnamefont {Queiroz}}, \bibinfo {author} {\bibfnamefont
  {T.}~\bibnamefont {Taniguchi}}, \bibinfo {author} {\bibfnamefont
  {K.}~\bibnamefont {Watanabe}}, \bibinfo {author} {\bibfnamefont
  {Y.}~\bibnamefont {Oreg}}, \bibinfo {author} {\bibfnamefont {F.}~\bibnamefont
  {von Oppen}}, \bibinfo {author} {\bibfnamefont {A.}~\bibnamefont {Stern}},
  \bibinfo {author} {\bibfnamefont {E.}~\bibnamefont {Berg}}, \bibinfo {author}
  {\bibfnamefont {P.}~\bibnamefont {Jarillo-Herrero}}, \ and\ \bibinfo {author}
  {\bibfnamefont {S.}~\bibnamefont {Ilani}},\ }\bibfield  {title} {\enquote
  {\bibinfo {title} {Cascade of phase transitions and dirac revivals in
  magic-angle graphene},}\ }\href {\doibase 10.1038/s41586-020-2373-y}
  {\bibfield  {journal} {\bibinfo  {journal} {Nature}\ }\textbf {\bibinfo
  {volume} {582}},\ \bibinfo {pages} {203--208} (\bibinfo {year}
  {2020})}\BibitemShut {NoStop}%
\bibitem [{\citenamefont {Turiansky}, \citenamefont {Alkauskas},\ and\
  \citenamefont {Van~de Walle}(2020)}]{Turiansky2020}%
  \BibitemOpen
  \bibfield  {author} {\bibinfo {author} {\bibfnamefont {M.~E.}\ \bibnamefont
  {Turiansky}}, \bibinfo {author} {\bibfnamefont {A.}~\bibnamefont
  {Alkauskas}}, \ and\ \bibinfo {author} {\bibfnamefont {C.~G.}\ \bibnamefont
  {Van~de Walle}},\ }\bibfield  {title} {\enquote {\bibinfo {title} {Spinning
  up quantum defects in 2d materials},}\ }\href {\doibase
  10.1038/s41563-020-0668-x} {\bibfield  {journal} {\bibinfo  {journal} {Nature
  Materials}\ }\textbf {\bibinfo {volume} {19}},\ \bibinfo {pages} {487--489}
  (\bibinfo {year} {2020})}\BibitemShut {NoStop}%
\bibitem [{\citenamefont {Gottscholl}\ \emph {et~al.}(2020)\citenamefont
  {Gottscholl}, \citenamefont {Kianinia}, \citenamefont {Soltamov},
  \citenamefont {Orlinskii}, \citenamefont {Mamin}, \citenamefont {Bradac},
  \citenamefont {Kasper}, \citenamefont {Krambrock}, \citenamefont {Sperlich},
  \citenamefont {Toth}, \citenamefont {Aharonovich},\ and\ \citenamefont
  {Dyakonov}}]{Gottscholl2020}%
  \BibitemOpen
  \bibfield  {author} {\bibinfo {author} {\bibfnamefont {A.}~\bibnamefont
  {Gottscholl}}, \bibinfo {author} {\bibfnamefont {M.}~\bibnamefont
  {Kianinia}}, \bibinfo {author} {\bibfnamefont {V.}~\bibnamefont {Soltamov}},
  \bibinfo {author} {\bibfnamefont {S.}~\bibnamefont {Orlinskii}}, \bibinfo
  {author} {\bibfnamefont {G.}~\bibnamefont {Mamin}}, \bibinfo {author}
  {\bibfnamefont {C.}~\bibnamefont {Bradac}}, \bibinfo {author} {\bibfnamefont
  {C.}~\bibnamefont {Kasper}}, \bibinfo {author} {\bibfnamefont
  {K.}~\bibnamefont {Krambrock}}, \bibinfo {author} {\bibfnamefont
  {A.}~\bibnamefont {Sperlich}}, \bibinfo {author} {\bibfnamefont
  {M.}~\bibnamefont {Toth}}, \bibinfo {author} {\bibfnamefont {I.}~\bibnamefont
  {Aharonovich}}, \ and\ \bibinfo {author} {\bibfnamefont {V.}~\bibnamefont
  {Dyakonov}},\ }\bibfield  {title} {\enquote {\bibinfo {title} {Initialization
  and read-out of intrinsic spin defects in a van der waals crystal at room
  temperature},}\ }\href {\doibase 10.1038/s41563-020-0619-6} {\bibfield
  {journal} {\bibinfo  {journal} {Nature Materials}\ }\textbf {\bibinfo
  {volume} {19}},\ \bibinfo {pages} {540--545} (\bibinfo {year}
  {2020})}\BibitemShut {NoStop}%
\bibitem [{\citenamefont {Wang}\ \emph {et~al.}(2015)\citenamefont {Wang},
  \citenamefont {Axline}, \citenamefont {Gao}, \citenamefont {Brecht},
  \citenamefont {Chu}, \citenamefont {Frunzio}, \citenamefont {Devoret},\ and\
  \citenamefont {Schoelkopf}}]{Wang_2015}%
  \BibitemOpen
  \bibfield  {author} {\bibinfo {author} {\bibfnamefont {C.}~\bibnamefont
  {Wang}}, \bibinfo {author} {\bibfnamefont {C.}~\bibnamefont {Axline}},
  \bibinfo {author} {\bibfnamefont {Y.~Y.}\ \bibnamefont {Gao}}, \bibinfo
  {author} {\bibfnamefont {T.}~\bibnamefont {Brecht}}, \bibinfo {author}
  {\bibfnamefont {Y.}~\bibnamefont {Chu}}, \bibinfo {author} {\bibfnamefont
  {L.}~\bibnamefont {Frunzio}}, \bibinfo {author} {\bibfnamefont {M.~H.}\
  \bibnamefont {Devoret}}, \ and\ \bibinfo {author} {\bibfnamefont {R.~J.}\
  \bibnamefont {Schoelkopf}},\ }\bibfield  {title} {\enquote {\bibinfo {title}
  {Surface participation and dielectric loss in superconducting qubits},}\
  }\href {\doibase 10.1063/1.4934486} {\bibfield  {journal} {\bibinfo
  {journal} {Applied Physics Letters}\ }\textbf {\bibinfo {volume} {107}},\
  \bibinfo {pages} {162601} (\bibinfo {year} {2015})}\BibitemShut {NoStop}%
\bibitem [{\citenamefont {Qiu}, \citenamefont {da~Jornada},\ and\ \citenamefont
  {Louie}(2017)}]{Qiu2017}%
  \BibitemOpen
  \bibfield  {author} {\bibinfo {author} {\bibfnamefont {D.~Y.}\ \bibnamefont
  {Qiu}}, \bibinfo {author} {\bibfnamefont {F.~H.}\ \bibnamefont {da~Jornada}},
  \ and\ \bibinfo {author} {\bibfnamefont {S.~G.}\ \bibnamefont {Louie}},\
  }\bibfield  {title} {\enquote {\bibinfo {title} {Environmental screening
  effects in 2d materials: Renormalization of the bandgap, electronic
  structure, and optical spectra of few-layer black phosphorus},}\ }\href
  {\doibase 10.1021/acs.nanolett.7b01365} {\bibfield  {journal} {\bibinfo
  {journal} {Nano Letters}\ }\textbf {\bibinfo {volume} {17}},\ \bibinfo
  {pages} {4706--4712} (\bibinfo {year} {2017})}\BibitemShut {NoStop}%
\bibitem [{\citenamefont {Tartakovskii}(2020)}]{Tartakovskii2020}%
  \BibitemOpen
  \bibfield  {author} {\bibinfo {author} {\bibfnamefont {A.}~\bibnamefont
  {Tartakovskii}},\ }\bibfield  {title} {\enquote {\bibinfo {title} {Moir{\'e}
  or not},}\ }\href {\doibase 10.1038/s41563-020-0693-9} {\bibfield  {journal}
  {\bibinfo  {journal} {Nature Materials}\ }\textbf {\bibinfo {volume} {19}},\
  \bibinfo {pages} {581--582} (\bibinfo {year} {2020})}\BibitemShut {NoStop}%
\bibitem [{\citenamefont {Yuan}\ \emph {et~al.}(2020)\citenamefont {Yuan},
  \citenamefont {Zheng}, \citenamefont {Kunstmann}, \citenamefont {Brumme},
  \citenamefont {Kuc}, \citenamefont {Ma}, \citenamefont {Deng}, \citenamefont
  {Blach}, \citenamefont {Pan},\ and\ \citenamefont {Huang}}]{Yuan2020}%
  \BibitemOpen
  \bibfield  {author} {\bibinfo {author} {\bibfnamefont {L.}~\bibnamefont
  {Yuan}}, \bibinfo {author} {\bibfnamefont {B.}~\bibnamefont {Zheng}},
  \bibinfo {author} {\bibfnamefont {J.}~\bibnamefont {Kunstmann}}, \bibinfo
  {author} {\bibfnamefont {T.}~\bibnamefont {Brumme}}, \bibinfo {author}
  {\bibfnamefont {A.~B.}\ \bibnamefont {Kuc}}, \bibinfo {author} {\bibfnamefont
  {C.}~\bibnamefont {Ma}}, \bibinfo {author} {\bibfnamefont {S.}~\bibnamefont
  {Deng}}, \bibinfo {author} {\bibfnamefont {D.}~\bibnamefont {Blach}},
  \bibinfo {author} {\bibfnamefont {A.}~\bibnamefont {Pan}}, \ and\ \bibinfo
  {author} {\bibfnamefont {L.}~\bibnamefont {Huang}},\ }\bibfield  {title}
  {\enquote {\bibinfo {title} {Twist-angle-dependent interlayer exciton
  diffusion in ws2--wse2 heterobilayers},}\ }\href {\doibase
  10.1038/s41563-020-0670-3} {\bibfield  {journal} {\bibinfo  {journal} {Nature
  Materials}\ }\textbf {\bibinfo {volume} {19}},\ \bibinfo {pages} {617--623}
  (\bibinfo {year} {2020})}\BibitemShut {NoStop}%
\bibitem [{\citenamefont {Martin}, \citenamefont {Reining},\ and\ \citenamefont
  {Ceperley}(2016)}]{martin2016interacting}%
  \BibitemOpen
  \bibfield  {author} {\bibinfo {author} {\bibfnamefont {R.~M.}\ \bibnamefont
  {Martin}}, \bibinfo {author} {\bibfnamefont {L.}~\bibnamefont {Reining}}, \
  and\ \bibinfo {author} {\bibfnamefont {D.~M.}\ \bibnamefont {Ceperley}},\
  }\href@noop {} {\emph {\bibinfo {title} {Interacting Electrons}}}\ (\bibinfo
  {publisher} {Cambridge University Press},\ \bibinfo {year}
  {2016})\BibitemShut {NoStop}%
\bibitem [{\citenamefont {Fetter}\ and\ \citenamefont
  {Walecka}(2003)}]{FetterWalecka}%
  \BibitemOpen
  \bibfield  {author} {\bibinfo {author} {\bibfnamefont {A.~L.}\ \bibnamefont
  {Fetter}}\ and\ \bibinfo {author} {\bibfnamefont {J.~D.}\ \bibnamefont
  {Walecka}},\ }\href@noop {} {\emph {\bibinfo {title} {Quantum Theory of
  Many-Particle Systems}}}\ (\bibinfo  {publisher} {Dover Publications},\
  \bibinfo {year} {2003})\BibitemShut {NoStop}%
\bibitem [{\citenamefont {Hedin}(1965)}]{Hedin1965}%
  \BibitemOpen
  \bibfield  {author} {\bibinfo {author} {\bibfnamefont {L.}~\bibnamefont
  {Hedin}},\ }\bibfield  {title} {\enquote {\bibinfo {title} {{New Method for
  Calculating the One-Particle Green's Function with Application to the
  Electron-Gas Problem}},}\ }\href {\doibase 10.1103/PhysRev.139.A796}
  {\bibfield  {journal} {\bibinfo  {journal} {Phys. Rev.}\ }\textbf {\bibinfo
  {volume} {139}},\ \bibinfo {pages} {A796--A823} (\bibinfo {year}
  {1965})}\BibitemShut {NoStop}%
\bibitem [{\citenamefont {Hybertsen}\ and\ \citenamefont
  {Louie}(1986)}]{Hybertsen_1986}%
  \BibitemOpen
  \bibfield  {author} {\bibinfo {author} {\bibfnamefont {M.~S.}\ \bibnamefont
  {Hybertsen}}\ and\ \bibinfo {author} {\bibfnamefont {S.~G.}\ \bibnamefont
  {Louie}},\ }\bibfield  {title} {\enquote {\bibinfo {title} {Electron
  correlation in semiconductors and insulators: Band gaps and quasiparticle
  energies},}\ }\href {\doibase 10.1103/PhysRevB.34.5390} {\bibfield  {journal}
  {\bibinfo  {journal} {Phys. Rev. B}\ }\textbf {\bibinfo {volume} {34}},\
  \bibinfo {pages} {5390--5413} (\bibinfo {year} {1986})}\BibitemShut {NoStop}%
\bibitem [{\citenamefont {Aryasetiawan}\ and\ \citenamefont
  {Gunnarsson}(1998)}]{Aryasetiawan1998}%
  \BibitemOpen
  \bibfield  {author} {\bibinfo {author} {\bibfnamefont {F.}~\bibnamefont
  {Aryasetiawan}}\ and\ \bibinfo {author} {\bibfnamefont {O.}~\bibnamefont
  {Gunnarsson}},\ }\bibfield  {title} {\enquote {\bibinfo
  {title} {{The GW method}}}}\href {\doibase 10.1088/0034-4885/61/3/002}
  {\bibfield  {journal} {\bibinfo  {journal} {Reports Prog. Phys.}\ }\textbf
  {\bibinfo {volume} {61}},\ \bibinfo {pages} {237--312} (\bibinfo {year}
  {1998})}\BibitemShut {NoStop}%
\bibitem [{\citenamefont {Govoni}\ and\ \citenamefont
  {Galli}(2015)}]{Govoni2015}%
  \BibitemOpen
  \bibfield  {author} {\bibinfo {author} {\bibfnamefont {M.}~\bibnamefont
  {Govoni}}\ and\ \bibinfo {author} {\bibfnamefont {G.}~\bibnamefont {Galli}},\
  }\bibfield  {title} {\enquote {\bibinfo {title} {Large scale gw
  calculations},}\ }\href {\doibase 10.1021/ct500958p} {\bibfield  {journal}
  {\bibinfo  {journal} {Journal of Chemical Theory and Computation}\ }\textbf
  {\bibinfo {volume} {11}},\ \bibinfo {pages} {2680--2696} (\bibinfo {year}
  {2015})}\BibitemShut {NoStop}%
\bibitem [{\citenamefont {Neuhauser}\ \emph {et~al.}(2014)\citenamefont
  {Neuhauser}, \citenamefont {Gao}, \citenamefont {Arntsen}, \citenamefont
  {Karshenas}, \citenamefont {Rabani},\ and\ \citenamefont
  {Baer}}]{neuhauser2014breaking}%
  \BibitemOpen
  \bibfield  {author} {\bibinfo {author} {\bibfnamefont {D.}~\bibnamefont
  {Neuhauser}}, \bibinfo {author} {\bibfnamefont {Y.}~\bibnamefont {Gao}},
  \bibinfo {author} {\bibfnamefont {C.}~\bibnamefont {Arntsen}}, \bibinfo
  {author} {\bibfnamefont {C.}~\bibnamefont {Karshenas}}, \bibinfo {author}
  {\bibfnamefont {E.}~\bibnamefont {Rabani}}, \ and\ \bibinfo {author}
  {\bibfnamefont {R.}~\bibnamefont {Baer}},\ }\bibfield  {title} {\enquote
  {\bibinfo {title} {{Breaking the Theoretical Scaling Limit for Predicting
  Quasiparticle Energies: The Stochastic G W Approach}},}\ }\href@noop {}
  {\bibfield  {journal} {\bibinfo  {journal} {Phys. Rev. Lett.}\ }\textbf
  {\bibinfo {volume} {113}},\ \bibinfo {pages} {076402} (\bibinfo {year}
  {2014})}\BibitemShut {NoStop}%
\bibitem [{\citenamefont {Vl\ifmmode~\check{c}\else \v{c}\fi{}ek}\ \emph
  {et~al.}(2018)\citenamefont {Vl\ifmmode~\check{c}\else \v{c}\fi{}ek},
  \citenamefont {Li}, \citenamefont {Baer}, \citenamefont {Rabani},\ and\
  \citenamefont {Neuhauser}}]{Vlcek2018swift}%
  \BibitemOpen
  \bibfield  {author} {\bibinfo {author} {\bibfnamefont {V.~c.~v.}\
  \bibnamefont {Vl\ifmmode~\check{c}\else \v{c}\fi{}ek}}, \bibinfo {author}
  {\bibfnamefont {W.}~\bibnamefont {Li}}, \bibinfo {author} {\bibfnamefont
  {R.}~\bibnamefont {Baer}}, \bibinfo {author} {\bibfnamefont {E.}~\bibnamefont
  {Rabani}}, \ and\ \bibinfo {author} {\bibfnamefont {D.}~\bibnamefont
  {Neuhauser}},\ }\bibfield  {title} {\enquote {\bibinfo {title} {Swift $gw$
  beyond 10,000 electrons using sparse stochastic compression},}\ }\href
  {\doibase 10.1103/PhysRevB.98.075107} {\bibfield  {journal} {\bibinfo
  {journal} {Phys. Rev. B}\ }\textbf {\bibinfo {volume} {98}},\ \bibinfo
  {pages} {075107} (\bibinfo {year} {2018})}\BibitemShut {NoStop}%
\bibitem [{\citenamefont {Vlcek}\ \emph {et~al.}(2017)\citenamefont {Vlcek},
  \citenamefont {Rabani}, \citenamefont {Neuhauser},\ and\ \citenamefont
  {Baer}}]{vlcek2017stochastic}%
  \BibitemOpen
  \bibfield  {author} {\bibinfo {author} {\bibfnamefont {V.}~\bibnamefont
  {Vlcek}}, \bibinfo {author} {\bibfnamefont {E.}~\bibnamefont {Rabani}},
  \bibinfo {author} {\bibfnamefont {D.}~\bibnamefont {Neuhauser}}, \ and\
  \bibinfo {author} {\bibfnamefont {R.}~\bibnamefont {Baer}},\ }\bibfield
  {title} {\enquote {\bibinfo {title} {Stochastic gw calculations for
  molecules},}\ }\href@noop {} {\bibfield  {journal} {\bibinfo  {journal} {J.
  Chem. Theory Comput.}\ }\textbf {\bibinfo {volume} {13}},\ \bibinfo {pages}
  {4997--5003} (\bibinfo {year} {2017})}\BibitemShut {NoStop}%
\bibitem [{\citenamefont {Brooks}\ \emph {et~al.}(2020)\citenamefont {Brooks},
  \citenamefont {Weng}, \citenamefont {Taylor},\ and\ \citenamefont
  {Vlcek}}]{Brooks_2020}%
  \BibitemOpen
  \bibfield  {author} {\bibinfo {author} {\bibfnamefont {J.}~\bibnamefont
  {Brooks}}, \bibinfo {author} {\bibfnamefont {G.}~\bibnamefont {Weng}},
  \bibinfo {author} {\bibfnamefont {S.}~\bibnamefont {Taylor}}, \ and\ \bibinfo
  {author} {\bibfnamefont {V.}~\bibnamefont {Vlcek}},\ }\bibfield  {title}
  {\enquote {\bibinfo {title} {Stochastic many-body perturbation theory for
  moir{\'{e}} states in twisted bilayer phosphorene},}\ }\href {\doibase
  10.1088/1361-648x/ab6d8c} {\bibfield  {journal} {\bibinfo  {journal} {Journal
  of Physics: Condensed Matter}\ }\textbf {\bibinfo {volume} {32}},\ \bibinfo
  {pages} {234001} (\bibinfo {year} {2020})}\BibitemShut {NoStop}%
\bibitem [{\citenamefont {Li}\ \emph {et~al.}(2019)\citenamefont {Li},
  \citenamefont {Xu}, \citenamefont {Mendelson}, \citenamefont {Kianinia},
  \citenamefont {Toth},\ and\ \citenamefont {Aharonovich}}]{Li_2019}%
  \BibitemOpen
  \bibfield  {author} {\bibinfo {author} {\bibfnamefont {C.}~\bibnamefont
  {Li}}, \bibinfo {author} {\bibfnamefont {Z.-Q.}\ \bibnamefont {Xu}}, \bibinfo
  {author} {\bibfnamefont {N.}~\bibnamefont {Mendelson}}, \bibinfo {author}
  {\bibfnamefont {M.}~\bibnamefont {Kianinia}}, \bibinfo {author}
  {\bibfnamefont {M.}~\bibnamefont {Toth}}, \ and\ \bibinfo {author}
  {\bibfnamefont {I.}~\bibnamefont {Aharonovich}},\ }\bibfield  {title}
  {\enquote {\bibinfo {title} {Purification of single-photon emission from hbn
  using post-processing treatments},}\ }\href
  {https://www.degruyter.com/view/journals/nanoph/8/11/article-p2049.xml}
  {\bibfield  {journal} {\bibinfo  {journal} {Nanophotonics}\ }\textbf
  {\bibinfo {volume} {8}},\ \bibinfo {pages} {2049 -- 2055} (\bibinfo {year}
  {2019})}\BibitemShut {NoStop}%
\bibitem [{\citenamefont {Tran}\ \emph {et~al.}(2016)\citenamefont {Tran},
  \citenamefont {Bray}, \citenamefont {Ford}, \citenamefont {Toth},\ and\
  \citenamefont {Aharonovich}}]{Tran_2016}%
  \BibitemOpen
  \bibfield  {author} {\bibinfo {author} {\bibfnamefont {T.~T.}\ \bibnamefont
  {Tran}}, \bibinfo {author} {\bibfnamefont {K.}~\bibnamefont {Bray}}, \bibinfo
  {author} {\bibfnamefont {M.~J.}\ \bibnamefont {Ford}}, \bibinfo {author}
  {\bibfnamefont {M.}~\bibnamefont {Toth}}, \ and\ \bibinfo {author}
  {\bibfnamefont {I.}~\bibnamefont {Aharonovich}},\ }\bibfield  {title}
  {\enquote {\bibinfo {title} {Quantum emission from hexagonal boron nitride
  monolayers},}\ }\href {\doibase 10.1038/nnano.2015.242} {\bibfield  {journal}
  {\bibinfo  {journal} {Nature Nanotechnology}\ }\textbf {\bibinfo {volume}
  {11}},\ \bibinfo {pages} {37--41} (\bibinfo {year} {2016})}\BibitemShut
  {NoStop}%
\bibitem [{\citenamefont {Exarhos}\ \emph {et~al.}(2017)\citenamefont
  {Exarhos}, \citenamefont {Hopper}, \citenamefont {Grote}, \citenamefont
  {Alkauskas},\ and\ \citenamefont {Bassett}}]{Exarhos_2017}%
  \BibitemOpen
  \bibfield  {author} {\bibinfo {author} {\bibfnamefont {A.~L.}\ \bibnamefont
  {Exarhos}}, \bibinfo {author} {\bibfnamefont {D.~A.}\ \bibnamefont {Hopper}},
  \bibinfo {author} {\bibfnamefont {R.~R.}\ \bibnamefont {Grote}}, \bibinfo
  {author} {\bibfnamefont {A.}~\bibnamefont {Alkauskas}}, \ and\ \bibinfo
  {author} {\bibfnamefont {L.~C.}\ \bibnamefont {Bassett}},\ }\bibfield
  {title} {\enquote {\bibinfo {title} {Optical signatures of quantum emitters
  in suspended hexagonal boron nitride},}\ }\href {\doibase
  10.1021/acsnano.7b00665} {\bibfield  {journal} {\bibinfo  {journal} {ACS
  Nano}\ }\textbf {\bibinfo {volume} {11}},\ \bibinfo {pages} {3328--3336}
  (\bibinfo {year} {2017})}\BibitemShut {NoStop}%
\bibitem [{\citenamefont {Mendelson}\ \emph {et~al.}(2019)\citenamefont
  {Mendelson}, \citenamefont {Xu}, \citenamefont {Tran}, \citenamefont
  {Kianinia}, \citenamefont {Scott}, \citenamefont {Bradac}, \citenamefont
  {Aharonovich},\ and\ \citenamefont {Toth}}]{Mendelson_2019}%
  \BibitemOpen
  \bibfield  {author} {\bibinfo {author} {\bibfnamefont {N.}~\bibnamefont
  {Mendelson}}, \bibinfo {author} {\bibfnamefont {Z.-Q.}\ \bibnamefont {Xu}},
  \bibinfo {author} {\bibfnamefont {T.~T.}\ \bibnamefont {Tran}}, \bibinfo
  {author} {\bibfnamefont {M.}~\bibnamefont {Kianinia}}, \bibinfo {author}
  {\bibfnamefont {J.}~\bibnamefont {Scott}}, \bibinfo {author} {\bibfnamefont
  {C.}~\bibnamefont {Bradac}}, \bibinfo {author} {\bibfnamefont
  {I.}~\bibnamefont {Aharonovich}}, \ and\ \bibinfo {author} {\bibfnamefont
  {M.}~\bibnamefont {Toth}},\ }\bibfield  {title} {\enquote {\bibinfo {title}
  {Engineering and tuning of quantum emitters in few-layer hexagonal boron
  nitride},}\ }\href {\doibase 10.1021/acsnano.8b08511} {\bibfield  {journal}
  {\bibinfo  {journal} {ACS Nano}\ }\textbf {\bibinfo {volume} {13}},\ \bibinfo
  {pages} {3132--3140} (\bibinfo {year} {2019})}\BibitemShut {NoStop}%
\bibitem [{\citenamefont {Maggio}\ and\ \citenamefont
  {Kresse}(2017)}]{Maggio2017}%
  \BibitemOpen
  \bibfield  {author} {\bibinfo {author} {\bibfnamefont {E.}~\bibnamefont
  {Maggio}}\ and\ \bibinfo {author} {\bibfnamefont {G.}~\bibnamefont
  {Kresse}},\ }\bibfield  {title} {\enquote {\bibinfo {title} {{GW} vertex
  corrected calculations for molecular systems},}\ }\href@noop {} {\bibfield
  {journal} {\bibinfo  {journal} {J. Chem. Theory Comput.}\ }\textbf {\bibinfo
  {volume} {13}},\ \bibinfo {pages} {4765} (\bibinfo {year}
  {2017})}\BibitemShut {NoStop}%
\bibitem [{\citenamefont {Hellgren}, \citenamefont {Colonna},\ and\
  \citenamefont {de~Gironcoli}(2018)}]{Hellgren2018}%
  \BibitemOpen
  \bibfield  {author} {\bibinfo {author} {\bibfnamefont {M.}~\bibnamefont
  {Hellgren}}, \bibinfo {author} {\bibfnamefont {N.}~\bibnamefont {Colonna}}, \
  and\ \bibinfo {author} {\bibfnamefont {S.}~\bibnamefont {de~Gironcoli}},\
  }\bibfield  {title} {\enquote {\bibinfo {title} {Beyond the random phase
  approximation with a local exchange vertex},}\ }\href@noop {} {\bibfield
  {journal} {\bibinfo  {journal} {Phys. Rev. A}\ }\textbf {\bibinfo {volume}
  {98}},\ \bibinfo {pages} {045117} (\bibinfo {year} {2018})}\BibitemShut
  {NoStop}%
\bibitem [{\citenamefont {Vlcek}(2019)}]{Vlcek2019vertex}%
  \BibitemOpen
  \bibfield  {author} {\bibinfo {author} {\bibfnamefont {V.}~\bibnamefont
  {Vlcek}},\ }\bibfield  {title} {\enquote {\bibinfo {title} {Stochastic vertex
  corrections: Linear scaling methods for accurate quasiparticle energies},}\
  }\href {\doibase 10.1021/acs.jctc.9b00317} {\bibfield  {journal} {\bibinfo
  {journal} {Journal of Chemical Theory and Computation}\ }\textbf {\bibinfo
  {volume} {15}},\ \bibinfo {pages} {6254--6266} (\bibinfo {year}
  {2019})}\BibitemShut {NoStop}%
\bibitem [{\citenamefont {Lewis}\ and\ \citenamefont
  {Berkelbach}(2019)}]{Lewis2019}%
  \BibitemOpen
  \bibfield  {author} {\bibinfo {author} {\bibfnamefont {A.~M.}\ \bibnamefont
  {Lewis}}\ and\ \bibinfo {author} {\bibfnamefont {T.~C.}\ \bibnamefont
  {Berkelbach}},\ }\bibfield  {title} {\enquote {\bibinfo {title} {Vertex
  corrections to the polarizability do not improve the gw approximation for the
  ionization potential of molecules},}\ }\href {\doibase
  10.1021/acs.jctc.8b00995} {\bibfield  {journal} {\bibinfo  {journal} {Journal
  of Chemical Theory and Computation}\ }\textbf {\bibinfo {volume} {15}},\
  \bibinfo {pages} {2925--2932} (\bibinfo {year} {2019})}\BibitemShut {NoStop}%
\bibitem [{\citenamefont {Baer}\ and\ \citenamefont
  {Neuhauser}(2004)}]{BaerNeuhauser2004}%
  \BibitemOpen
  \bibfield  {author} {\bibinfo {author} {\bibfnamefont {R.}~\bibnamefont
  {Baer}}\ and\ \bibinfo {author} {\bibfnamefont {D.}~\bibnamefont
  {Neuhauser}},\ }\bibfield  {title} {\enquote {\bibinfo {title} {Real-time
  linear response for time-dependent density-functional theory},}\ }\href@noop
  {} {\bibfield  {journal} {\bibinfo  {journal} {The Journal of chemical
  physics}\ }\textbf {\bibinfo {volume} {121}},\ \bibinfo {pages} {9803--9807}
  (\bibinfo {year} {2004})}\BibitemShut {NoStop}%
\bibitem [{\citenamefont {Gao}\ \emph {et~al.}(2015)\citenamefont {Gao},
  \citenamefont {Neuhauser}, \citenamefont {Baer},\ and\ \citenamefont
  {Rabani}}]{Gao_2015}%
  \BibitemOpen
  \bibfield  {author} {\bibinfo {author} {\bibfnamefont {Y.}~\bibnamefont
  {Gao}}, \bibinfo {author} {\bibfnamefont {D.}~\bibnamefont {Neuhauser}},
  \bibinfo {author} {\bibfnamefont {R.}~\bibnamefont {Baer}}, \ and\ \bibinfo
  {author} {\bibfnamefont {E.}~\bibnamefont {Rabani}},\ }\bibfield  {title}
  {\enquote {\bibinfo {title} {Sublinear scaling for time-dependent stochastic
  density functional theory},}\ }\href {\doibase 10.1063/1.4905568} {\bibfield
  {journal} {\bibinfo  {journal} {The Journal of Chemical Physics}\ }\textbf
  {\bibinfo {volume} {142}},\ \bibinfo {pages} {034106} (\bibinfo {year}
  {2015})}\BibitemShut {NoStop}%
\bibitem [{\citenamefont {Neuhauser}\ \emph {et~al.}(2016)\citenamefont
  {Neuhauser}, \citenamefont {Rabani}, \citenamefont {Cytter},\ and\
  \citenamefont {Baer}}]{Neuhauser_2016}%
  \BibitemOpen
  \bibfield  {author} {\bibinfo {author} {\bibfnamefont {D.}~\bibnamefont
  {Neuhauser}}, \bibinfo {author} {\bibfnamefont {E.}~\bibnamefont {Rabani}},
  \bibinfo {author} {\bibfnamefont {Y.}~\bibnamefont {Cytter}}, \ and\ \bibinfo
  {author} {\bibfnamefont {R.}~\bibnamefont {Baer}},\ }\bibfield  {title}
  {\enquote {\bibinfo {title} {Stochastic optimally tuned range-separated
  hybrid density functional theory},}\ }\href {\doibase
  10.1021/acs.jpca.5b10573} {\bibfield  {journal} {\bibinfo  {journal} {The
  Journal of Physical Chemistry A}\ }\textbf {\bibinfo {volume} {120}},\
  \bibinfo {pages} {3071--3078} (\bibinfo {year} {2016})}\BibitemShut {NoStop}%
\bibitem [{\citenamefont {Rabani}, \citenamefont {Baer},\ and\ \citenamefont
  {Neuhauser}(2015)}]{Rabani2015}%
  \BibitemOpen
  \bibfield  {author} {\bibinfo {author} {\bibfnamefont {E.}~\bibnamefont
  {Rabani}}, \bibinfo {author} {\bibfnamefont {R.}~\bibnamefont {Baer}}, \ and\
  \bibinfo {author} {\bibfnamefont {D.}~\bibnamefont {Neuhauser}},\ }\bibfield
  {title} {\enquote {\bibinfo {title} {{Time-dependent stochastic
  Bethe-Salpeter approach}},}\ }\href {\doibase 10.1103/PhysRevB.91.235302}
  {\bibfield  {journal} {\bibinfo  {journal} {Phys. Rev. B}\ }\textbf {\bibinfo
  {volume} {91}},\ \bibinfo {pages} {235302} (\bibinfo {year}
  {2015})}\BibitemShut {NoStop}%
\bibitem [{\citenamefont {Baroni}, \citenamefont {de~Gironcoli},\ and\
  \citenamefont {{Dal Corso}}(2001)}]{Baroni2001}%
  \BibitemOpen
  \bibfield  {author} {\bibinfo {author} {\bibfnamefont {S.}~\bibnamefont
  {Baroni}}, \bibinfo {author} {\bibfnamefont {S.}~\bibnamefont
  {de~Gironcoli}}, \ and\ \bibinfo {author} {\bibfnamefont {A.}~\bibnamefont
  {{Dal Corso}}},\ }\bibfield  {title} {\enquote {\bibinfo {title} {{Phonons
  and related crystal properties from density-functional perturbation
  theory}},}\ }\href {\doibase 10.1103/RevModPhys.73.515} {\bibfield  {journal}
  {\bibinfo  {journal} {Rev. Mod. Phys.}\ }\textbf {\bibinfo {volume} {73}},\
  \bibinfo {pages} {515--562} (\bibinfo {year} {2001})}\BibitemShut {NoStop}%
\bibitem [{\citenamefont {Neuhauser}\ and\ \citenamefont
  {Baer}(2005)}]{Neuhauser2005}%
  \BibitemOpen
  \bibfield  {author} {\bibinfo {author} {\bibfnamefont {D.}~\bibnamefont
  {Neuhauser}}\ and\ \bibinfo {author} {\bibfnamefont {R.}~\bibnamefont
  {Baer}},\ }\bibfield  {title} {\enquote {\bibinfo {title} {{Efficient
  linear-response method circumventing the exchange-correlation kernel: theory
  for molecular conductance under finite bias.}}}\ }\href {\doibase
  10.1063/1.2121607} {\bibfield  {journal} {\bibinfo  {journal} {J. Chem.
  Phys.}\ }\textbf {\bibinfo {volume} {123}},\ \bibinfo {pages} {204105}
  (\bibinfo {year} {2005})}\BibitemShut {NoStop}%
\bibitem [{\citenamefont {Troullier}\ and\ \citenamefont
  {Martins}(1991)}]{TroullierMartins1991}%
  \BibitemOpen
  \bibfield  {author} {\bibinfo {author} {\bibfnamefont {N.}~\bibnamefont
  {Troullier}}\ and\ \bibinfo {author} {\bibfnamefont {J.~L.}\ \bibnamefont
  {Martins}},\ }\bibfield  {title} {\enquote {\bibinfo {title} {Efficient
  pseudopotentials for plane-wave calculations},}\ }\href@noop {} {\bibfield
  {journal} {\bibinfo  {journal} {Phys. Rev. B}\ }\textbf {\bibinfo {volume}
  {43}},\ \bibinfo {pages} {1993} (\bibinfo {year} {1991})}\BibitemShut
  {NoStop}%
\bibitem [{\citenamefont {Perdew}\ and\ \citenamefont
  {Wang}(1992)}]{PerdewWang}%
  \BibitemOpen
  \bibfield  {author} {\bibinfo {author} {\bibfnamefont {J.~P.}\ \bibnamefont
  {Perdew}}\ and\ \bibinfo {author} {\bibfnamefont {Y.}~\bibnamefont {Wang}},\
  }\bibfield  {title} {\enquote {\bibinfo {title} {{Accurate and simple
  analytic representation of the electron-gas correlation energy}},}\ }\href
  {\doibase 10.1103/PhysRevB.45.13244} {\bibfield  {journal} {\bibinfo
  {journal} {Phys. Rev. B}\ }\textbf {\bibinfo {volume} {45}},\ \bibinfo
  {pages} {13244--13249} (\bibinfo {year} {1992})}\BibitemShut {NoStop}%
\bibitem [{\citenamefont {Rozzi}\ \emph {et~al.}(2006)\citenamefont {Rozzi},
  \citenamefont {Varsano}, \citenamefont {Marini}, \citenamefont {Gross},\ and\
  \citenamefont {Rubio}}]{Rozzi_2006}%
  \BibitemOpen
  \bibfield  {author} {\bibinfo {author} {\bibfnamefont {C.~A.}\ \bibnamefont
  {Rozzi}}, \bibinfo {author} {\bibfnamefont {D.}~\bibnamefont {Varsano}},
  \bibinfo {author} {\bibfnamefont {A.}~\bibnamefont {Marini}}, \bibinfo
  {author} {\bibfnamefont {E.~K.~U.}\ \bibnamefont {Gross}}, \ and\ \bibinfo
  {author} {\bibfnamefont {A.}~\bibnamefont {Rubio}},\ }\bibfield  {title}
  {\enquote {\bibinfo {title} {Exact coulomb cutoff technique for supercell
  calculations},}\ }\href {\doibase 10.1103/PhysRevB.73.205119} {\bibfield
  {journal} {\bibinfo  {journal} {Phys. Rev. B}\ }\textbf {\bibinfo {volume}
  {73}},\ \bibinfo {pages} {205119} (\bibinfo {year} {2006})}\BibitemShut
  {NoStop}%
\bibitem [{\citenamefont {Giannozzi}\ \emph {et~al.}(2017)\citenamefont
  {Giannozzi}, \citenamefont {Andreussi}, \citenamefont {Brumme}, \citenamefont
  {Bunau}, \citenamefont {Nardelli}, \citenamefont {Calandra}, \citenamefont
  {Car}, \citenamefont {Cavazzoni}, \citenamefont {Ceresoli}, \citenamefont
  {Cococcioni}, \citenamefont {Colonna}, \citenamefont {Carnimeo},
  \citenamefont {Corso}, \citenamefont {de~Gironcoli}, \citenamefont {Delugas},
  \citenamefont {Jr}, \citenamefont {Ferretti}, \citenamefont {Floris},
  \citenamefont {Fratesi}, \citenamefont {Fugallo}, \citenamefont {Gebauer},
  \citenamefont {Gerstmann}, \citenamefont {Giustino}, \citenamefont {Gorni},
  \citenamefont {Jia}, \citenamefont {Kawamura}, \citenamefont {Ko},
  \citenamefont {Kokalj}, \citenamefont {Küçükbenli}, \citenamefont
  {Lazzeri}, \citenamefont {Marsili}, \citenamefont {Marzari}, \citenamefont
  {Mauri}, \citenamefont {Nguyen}, \citenamefont {Nguyen}, \citenamefont {de-la
  Roza}, \citenamefont {Paulatto}, \citenamefont {Poncé}, \citenamefont
  {Rocca}, \citenamefont {Sabatini}, \citenamefont {Santra}, \citenamefont
  {Schlipf}, \citenamefont {Seitsonen}, \citenamefont {Smogunov}, \citenamefont
  {Timrov}, \citenamefont {Thonhauser}, \citenamefont {Umari}, \citenamefont
  {Vast}, \citenamefont {Wu},\ and\ \citenamefont {Baroni}}]{QE2017}%
  \BibitemOpen
  \bibfield  {author} {\bibinfo {author} {\bibfnamefont {P.}~\bibnamefont
  {Giannozzi}}, \bibinfo {author} {\bibfnamefont {O.}~\bibnamefont
  {Andreussi}}, \bibinfo {author} {\bibfnamefont {T.}~\bibnamefont {Brumme}},
  \bibinfo {author} {\bibfnamefont {O.}~\bibnamefont {Bunau}}, \bibinfo
  {author} {\bibfnamefont {M.~B.}\ \bibnamefont {Nardelli}}, \bibinfo {author}
  {\bibfnamefont {M.}~\bibnamefont {Calandra}}, \bibinfo {author}
  {\bibfnamefont {R.}~\bibnamefont {Car}}, \bibinfo {author} {\bibfnamefont
  {C.}~\bibnamefont {Cavazzoni}}, \bibinfo {author} {\bibfnamefont
  {D.}~\bibnamefont {Ceresoli}}, \bibinfo {author} {\bibfnamefont
  {M.}~\bibnamefont {Cococcioni}}, \bibinfo {author} {\bibfnamefont
  {N.}~\bibnamefont {Colonna}}, \bibinfo {author} {\bibfnamefont
  {I.}~\bibnamefont {Carnimeo}}, \bibinfo {author} {\bibfnamefont {A.~D.}\
  \bibnamefont {Corso}}, \bibinfo {author} {\bibfnamefont {S.}~\bibnamefont
  {de~Gironcoli}}, \bibinfo {author} {\bibfnamefont {P.}~\bibnamefont
  {Delugas}}, \bibinfo {author} {\bibfnamefont {R.~A.~D.}\ \bibnamefont {Jr}},
  \bibinfo {author} {\bibfnamefont {A.}~\bibnamefont {Ferretti}}, \bibinfo
  {author} {\bibfnamefont {A.}~\bibnamefont {Floris}}, \bibinfo {author}
  {\bibfnamefont {G.}~\bibnamefont {Fratesi}}, \bibinfo {author} {\bibfnamefont
  {G.}~\bibnamefont {Fugallo}}, \bibinfo {author} {\bibfnamefont
  {R.}~\bibnamefont {Gebauer}}, \bibinfo {author} {\bibfnamefont
  {U.}~\bibnamefont {Gerstmann}}, \bibinfo {author} {\bibfnamefont
  {F.}~\bibnamefont {Giustino}}, \bibinfo {author} {\bibfnamefont
  {T.}~\bibnamefont {Gorni}}, \bibinfo {author} {\bibfnamefont
  {J.}~\bibnamefont {Jia}}, \bibinfo {author} {\bibfnamefont {M.}~\bibnamefont
  {Kawamura}}, \bibinfo {author} {\bibfnamefont {H.-Y.}\ \bibnamefont {Ko}},
  \bibinfo {author} {\bibfnamefont {A.}~\bibnamefont {Kokalj}}, \bibinfo
  {author} {\bibfnamefont {E.}~\bibnamefont {Küçükbenli}}, \bibinfo {author}
  {\bibfnamefont {M.}~\bibnamefont {Lazzeri}}, \bibinfo {author} {\bibfnamefont
  {M.}~\bibnamefont {Marsili}}, \bibinfo {author} {\bibfnamefont
  {N.}~\bibnamefont {Marzari}}, \bibinfo {author} {\bibfnamefont
  {F.}~\bibnamefont {Mauri}}, \bibinfo {author} {\bibfnamefont {N.~L.}\
  \bibnamefont {Nguyen}}, \bibinfo {author} {\bibfnamefont {H.-V.}\
  \bibnamefont {Nguyen}}, \bibinfo {author} {\bibfnamefont {A.~O.}\
  \bibnamefont {de-la Roza}}, \bibinfo {author} {\bibfnamefont
  {L.}~\bibnamefont {Paulatto}}, \bibinfo {author} {\bibfnamefont
  {S.}~\bibnamefont {Poncé}}, \bibinfo {author} {\bibfnamefont
  {D.}~\bibnamefont {Rocca}}, \bibinfo {author} {\bibfnamefont
  {R.}~\bibnamefont {Sabatini}}, \bibinfo {author} {\bibfnamefont
  {B.}~\bibnamefont {Santra}}, \bibinfo {author} {\bibfnamefont
  {M.}~\bibnamefont {Schlipf}}, \bibinfo {author} {\bibfnamefont {A.~P.}\
  \bibnamefont {Seitsonen}}, \bibinfo {author} {\bibfnamefont {A.}~\bibnamefont
  {Smogunov}}, \bibinfo {author} {\bibfnamefont {I.}~\bibnamefont {Timrov}},
  \bibinfo {author} {\bibfnamefont {T.}~\bibnamefont {Thonhauser}}, \bibinfo
  {author} {\bibfnamefont {P.}~\bibnamefont {Umari}}, \bibinfo {author}
  {\bibfnamefont {N.}~\bibnamefont {Vast}}, \bibinfo {author} {\bibfnamefont
  {X.}~\bibnamefont {Wu}}, \ and\ \bibinfo {author} {\bibfnamefont
  {S.}~\bibnamefont {Baroni}},\ }\bibfield  {title} {\enquote {\bibinfo {title}
  {Advanced capabilities for materials modelling with quantum espresso},}\
  }\href {http://stacks.iop.org/0953-8984/29/i=46/a=465901} {\bibfield
  {journal} {\bibinfo  {journal} {Journal of Physics: Condensed Matter}\
  }\textbf {\bibinfo {volume} {29}},\ \bibinfo {pages} {465901} (\bibinfo
  {year} {2017})}\BibitemShut {NoStop}%
\bibitem [{\citenamefont {Tkatchenko}\ and\ \citenamefont
  {Scheffler}(2009)}]{TS_2009}%
  \BibitemOpen
  \bibfield  {author} {\bibinfo {author} {\bibfnamefont {A.}~\bibnamefont
  {Tkatchenko}}\ and\ \bibinfo {author} {\bibfnamefont {M.}~\bibnamefont
  {Scheffler}},\ }\bibfield  {title} {\enquote {\bibinfo {title} {Accurate
  molecular van der waals interactions from ground-state electron density and
  free-atom reference data},}\ }\href {\doibase 10.1103/PhysRevLett.102.073005}
  {\bibfield  {journal} {\bibinfo  {journal} {Phys. Rev. Lett.}\ }\textbf
  {\bibinfo {volume} {102}},\ \bibinfo {pages} {073005} (\bibinfo {year}
  {2009})}\BibitemShut {NoStop}%
\bibitem [{\citenamefont {Park}, \citenamefont {Park},\ and\ \citenamefont
  {Kim}(2014)}]{Park_2014}%
  \BibitemOpen
  \bibfield  {author} {\bibinfo {author} {\bibfnamefont {S.}~\bibnamefont
  {Park}}, \bibinfo {author} {\bibfnamefont {C.}~\bibnamefont {Park}}, \ and\
  \bibinfo {author} {\bibfnamefont {G.}~\bibnamefont {Kim}},\ }\bibfield
  {title} {\enquote {\bibinfo {title} {Interlayer coupling enhancement in
  graphene/hexagonal boron nitride heterostructures by intercalated defects or
  vacancies},}\ }\href {\doibase 10.1063/1.4870097} {\bibfield  {journal}
  {\bibinfo  {journal} {The Journal of Chemical Physics}\ }\textbf {\bibinfo
  {volume} {140}},\ \bibinfo {pages} {134706} (\bibinfo {year}
  {2014})}\BibitemShut {NoStop}%
\bibitem [{\citenamefont {Attaccalite}\ \emph {et~al.}(2011)\citenamefont
  {Attaccalite}, \citenamefont {Bockstedte}, \citenamefont {Marini},
  \citenamefont {Rubio},\ and\ \citenamefont {Wirtz}}]{Attaccalite_2011}%
  \BibitemOpen
  \bibfield  {author} {\bibinfo {author} {\bibfnamefont {C.}~\bibnamefont
  {Attaccalite}}, \bibinfo {author} {\bibfnamefont {M.}~\bibnamefont
  {Bockstedte}}, \bibinfo {author} {\bibfnamefont {A.}~\bibnamefont {Marini}},
  \bibinfo {author} {\bibfnamefont {A.}~\bibnamefont {Rubio}}, \ and\ \bibinfo
  {author} {\bibfnamefont {L.}~\bibnamefont {Wirtz}},\ }\bibfield  {title}
  {\enquote {\bibinfo {title} {Coupling of excitons and defect states in
  boron-nitride nanostructures},}\ }\href {\doibase 10.1103/PhysRevB.83.144115}
  {\bibfield  {journal} {\bibinfo  {journal} {Phys. Rev. B}\ }\textbf {\bibinfo
  {volume} {83}},\ \bibinfo {pages} {144115} (\bibinfo {year}
  {2011})}\BibitemShut {NoStop}%
\bibitem [{\citenamefont {Huang}\ and\ \citenamefont {Lee}(2012)}]{Huang_2012}%
  \BibitemOpen
  \bibfield  {author} {\bibinfo {author} {\bibfnamefont {B.}~\bibnamefont
  {Huang}}\ and\ \bibinfo {author} {\bibfnamefont {H.}~\bibnamefont {Lee}},\
  }\bibfield  {title} {\enquote {\bibinfo {title} {Defect and impurity
  properties of hexagonal boron nitride: A first-principles calculation},}\
  }\href {\doibase 10.1103/PhysRevB.86.245406} {\bibfield  {journal} {\bibinfo
  {journal} {Phys. Rev. B}\ }\textbf {\bibinfo {volume} {86}},\ \bibinfo
  {pages} {245406} (\bibinfo {year} {2012})}\BibitemShut {NoStop}%
\bibitem [{\citenamefont {McDougall}\ \emph {et~al.}(2017)\citenamefont
  {McDougall}, \citenamefont {Partridge}, \citenamefont {Nicholls},
  \citenamefont {Russo},\ and\ \citenamefont {McCulloch}}]{McDougall_2017}%
  \BibitemOpen
  \bibfield  {author} {\bibinfo {author} {\bibfnamefont {N.~L.}\ \bibnamefont
  {McDougall}}, \bibinfo {author} {\bibfnamefont {J.~G.}\ \bibnamefont
  {Partridge}}, \bibinfo {author} {\bibfnamefont {R.~J.}\ \bibnamefont
  {Nicholls}}, \bibinfo {author} {\bibfnamefont {S.~P.}\ \bibnamefont {Russo}},
  \ and\ \bibinfo {author} {\bibfnamefont {D.~G.}\ \bibnamefont {McCulloch}},\
  }\bibfield  {title} {\enquote {\bibinfo {title} {Influence of point defects
  on the near edge structure of hexagonal boron nitride},}\ }\href {\doibase
  10.1103/PhysRevB.96.144106} {\bibfield  {journal} {\bibinfo  {journal} {Phys.
  Rev. B}\ }\textbf {\bibinfo {volume} {96}},\ \bibinfo {pages} {144106}
  (\bibinfo {year} {2017})}\BibitemShut {NoStop}%
\bibitem [{\citenamefont {Levinshtein}, \citenamefont {Rumyantsev},\ and\
  \citenamefont {Shur}(2001)}]{Levinshtein_2001}%
  \BibitemOpen
  \bibfield  {author} {\bibinfo {author} {\bibfnamefont {M.~E.}\ \bibnamefont
  {Levinshtein}}, \bibinfo {author} {\bibfnamefont {S.~L.}\ \bibnamefont
  {Rumyantsev}}, \ and\ \bibinfo {author} {\bibfnamefont {M.~S.}\ \bibnamefont
  {Shur}},\ }\bibfield  {title} {\enquote {\bibinfo {title} {Properties of
  advanced semiconductor materials : Gan, aln, inn, bn, sic, sige},}\ \
  }(\bibinfo {year} {2001})\BibitemShut {NoStop}%
\bibitem [{\citenamefont {Fuchs}\ \emph {et~al.}(2007)\citenamefont {Fuchs},
  \citenamefont {Furthm\"uller}, \citenamefont {Bechstedt}, \citenamefont
  {Shishkin},\ and\ \citenamefont {Kresse}}]{Fuchs_2007}%
  \BibitemOpen
  \bibfield  {author} {\bibinfo {author} {\bibfnamefont {F.}~\bibnamefont
  {Fuchs}}, \bibinfo {author} {\bibfnamefont {J.}~\bibnamefont
  {Furthm\"uller}}, \bibinfo {author} {\bibfnamefont {F.}~\bibnamefont
  {Bechstedt}}, \bibinfo {author} {\bibfnamefont {M.}~\bibnamefont {Shishkin}},
  \ and\ \bibinfo {author} {\bibfnamefont {G.}~\bibnamefont {Kresse}},\
  }\bibfield  {title} {\enquote {\bibinfo {title} {Quasiparticle band structure
  based on a generalized kohn-sham scheme},}\ }\href {\doibase
  10.1103/PhysRevB.76.115109} {\bibfield  {journal} {\bibinfo  {journal} {Phys.
  Rev. B}\ }\textbf {\bibinfo {volume} {76}},\ \bibinfo {pages} {115109}
  (\bibinfo {year} {2007})}\BibitemShut {NoStop}%
\bibitem [{\citenamefont {H\"user}, \citenamefont {Olsen},\ and\ \citenamefont
  {Thygesen}(2013)}]{Huser_2013}%
  \BibitemOpen
  \bibfield  {author} {\bibinfo {author} {\bibfnamefont {F.}~\bibnamefont
  {H\"user}}, \bibinfo {author} {\bibfnamefont {T.}~\bibnamefont {Olsen}}, \
  and\ \bibinfo {author} {\bibfnamefont {K.~S.}\ \bibnamefont {Thygesen}},\
  }\bibfield  {title} {\enquote {\bibinfo {title} {Quasiparticle gw
  calculations for solids, molecules, and two-dimensional materials},}\ }\href
  {\doibase 10.1103/PhysRevB.87.235132} {\bibfield  {journal} {\bibinfo
  {journal} {Phys. Rev. B}\ }\textbf {\bibinfo {volume} {87}},\ \bibinfo
  {pages} {235132} (\bibinfo {year} {2013})}\BibitemShut {NoStop}%
\bibitem [{\citenamefont {Popescu}\ and\ \citenamefont
  {Zunger}(2012)}]{Popescu_2012}%
  \BibitemOpen
  \bibfield  {author} {\bibinfo {author} {\bibfnamefont {V.}~\bibnamefont
  {Popescu}}\ and\ \bibinfo {author} {\bibfnamefont {A.}~\bibnamefont
  {Zunger}},\ }\bibfield  {title} {\enquote {\bibinfo {title} {Extracting $e$
  versus $p\vec{k}$ effective band structure from supercell calculations on
  alloys and impurities},}\ }\href {\doibase 10.1103/PhysRevB.85.085201}
  {\bibfield  {journal} {\bibinfo  {journal} {Phys. Rev. B}\ }\textbf {\bibinfo
  {volume} {85}},\ \bibinfo {pages} {085201} (\bibinfo {year}
  {2012})}\BibitemShut {NoStop}%
\bibitem [{\citenamefont {Huang}\ \emph {et~al.}(2014)\citenamefont {Huang},
  \citenamefont {Zheng}, \citenamefont {Zhang}, \citenamefont {Wu},
  \citenamefont {Gu},\ and\ \citenamefont {Duan}}]{Huang_2014}%
  \BibitemOpen
  \bibfield  {author} {\bibinfo {author} {\bibfnamefont {H.}~\bibnamefont
  {Huang}}, \bibinfo {author} {\bibfnamefont {F.}~\bibnamefont {Zheng}},
  \bibinfo {author} {\bibfnamefont {P.}~\bibnamefont {Zhang}}, \bibinfo
  {author} {\bibfnamefont {J.}~\bibnamefont {Wu}}, \bibinfo {author}
  {\bibfnamefont {B.-L.}\ \bibnamefont {Gu}}, \ and\ \bibinfo {author}
  {\bibfnamefont {W.}~\bibnamefont {Duan}},\ }\bibfield  {title} {\enquote
  {\bibinfo {title} {A general group theoretical method to unfold band
  structures and its application},}\ }\href {\doibase
  10.1088/1367-2630/16/3/033034} {\bibfield  {journal} {\bibinfo  {journal}
  {New Journal of Physics}\ }\textbf {\bibinfo {volume} {16}},\ \bibinfo
  {pages} {033034} (\bibinfo {year} {2014})}\BibitemShut {NoStop}%
\bibitem [{\citenamefont {Medeiros}, \citenamefont {Stafstr\"om},\ and\
  \citenamefont {Bj\"ork}(2014)}]{Medeiros_2014}%
  \BibitemOpen
  \bibfield  {author} {\bibinfo {author} {\bibfnamefont {P.~V.~C.}\
  \bibnamefont {Medeiros}}, \bibinfo {author} {\bibfnamefont {S.}~\bibnamefont
  {Stafstr\"om}}, \ and\ \bibinfo {author} {\bibfnamefont {J.}~\bibnamefont
  {Bj\"ork}},\ }\bibfield  {title} {\enquote {\bibinfo {title} {Effects of
  extrinsic and intrinsic perturbations on the electronic structure of
  graphene: Retaining an effective primitive cell band structure by band
  unfolding},}\ }\href {\doibase 10.1103/PhysRevB.89.041407} {\bibfield
  {journal} {\bibinfo  {journal} {Phys. Rev. B}\ }\textbf {\bibinfo {volume}
  {89}},\ \bibinfo {pages} {041407} (\bibinfo {year} {2014})}\BibitemShut
  {NoStop}%
\bibitem [{\citenamefont {Cudazzo}\ \emph {et~al.}(2016)\citenamefont
  {Cudazzo}, \citenamefont {Sponza}, \citenamefont {Giorgetti}, \citenamefont
  {Reining}, \citenamefont {Sottile},\ and\ \citenamefont
  {Gatti}}]{Cudazzo_2016}%
  \BibitemOpen
  \bibfield  {author} {\bibinfo {author} {\bibfnamefont {P.}~\bibnamefont
  {Cudazzo}}, \bibinfo {author} {\bibfnamefont {L.}~\bibnamefont {Sponza}},
  \bibinfo {author} {\bibfnamefont {C.}~\bibnamefont {Giorgetti}}, \bibinfo
  {author} {\bibfnamefont {L.}~\bibnamefont {Reining}}, \bibinfo {author}
  {\bibfnamefont {F.}~\bibnamefont {Sottile}}, \ and\ \bibinfo {author}
  {\bibfnamefont {M.}~\bibnamefont {Gatti}},\ }\bibfield  {title} {\enquote
  {\bibinfo {title} {Exciton band structure in two-dimensional materials},}\
  }\href {\doibase 10.1103/PhysRevLett.116.066803} {\bibfield  {journal}
  {\bibinfo  {journal} {Phys. Rev. Lett.}\ }\textbf {\bibinfo {volume} {116}},\
  \bibinfo {pages} {066803} (\bibinfo {year} {2016})}\BibitemShut {NoStop}%
\bibitem [{\citenamefont {Paleari}\ \emph {et~al.}(2018)\citenamefont
  {Paleari}, \citenamefont {Galvani}, \citenamefont {Amara}, \citenamefont
  {Ducastelle}, \citenamefont {Molina-S{\'{a}}nchez},\ and\ \citenamefont
  {Wirtz}}]{Paleari_2018}%
  \BibitemOpen
  \bibfield  {author} {\bibinfo {author} {\bibfnamefont {F.}~\bibnamefont
  {Paleari}}, \bibinfo {author} {\bibfnamefont {T.}~\bibnamefont {Galvani}},
  \bibinfo {author} {\bibfnamefont {H.}~\bibnamefont {Amara}}, \bibinfo
  {author} {\bibfnamefont {F.}~\bibnamefont {Ducastelle}}, \bibinfo {author}
  {\bibfnamefont {A.}~\bibnamefont {Molina-S{\'{a}}nchez}}, \ and\ \bibinfo
  {author} {\bibfnamefont {L.}~\bibnamefont {Wirtz}},\ }\bibfield  {title}
  {\enquote {\bibinfo {title} {Excitons in few-layer hexagonal boron nitride:
  Davydov splitting and surface localization},}\ }\href {\doibase
  10.1088/2053-1583/aad586} {\bibfield  {journal} {\bibinfo  {journal} {2D
  Materials}\ }\textbf {\bibinfo {volume} {5}},\ \bibinfo {pages} {045017}
  (\bibinfo {year} {2018})}\BibitemShut {NoStop}%
\bibitem [{\citenamefont {Xu}\ \emph {et~al.}(2020)\citenamefont {Xu},
  \citenamefont {Mendelson}, \citenamefont {Scott}, \citenamefont {Li},
  \citenamefont {Abidi}, \citenamefont {Liu}, \citenamefont {Luo},
  \citenamefont {Aharonovich},\ and\ \citenamefont {Toth}}]{Xu_2020}%
  \BibitemOpen
  \bibfield  {author} {\bibinfo {author} {\bibfnamefont {Z.-Q.}\ \bibnamefont
  {Xu}}, \bibinfo {author} {\bibfnamefont {N.}~\bibnamefont {Mendelson}},
  \bibinfo {author} {\bibfnamefont {J.~A.}\ \bibnamefont {Scott}}, \bibinfo
  {author} {\bibfnamefont {C.}~\bibnamefont {Li}}, \bibinfo {author}
  {\bibfnamefont {I.~H.}\ \bibnamefont {Abidi}}, \bibinfo {author}
  {\bibfnamefont {H.}~\bibnamefont {Liu}}, \bibinfo {author} {\bibfnamefont
  {Z.}~\bibnamefont {Luo}}, \bibinfo {author} {\bibfnamefont {I.}~\bibnamefont
  {Aharonovich}}, \ and\ \bibinfo {author} {\bibfnamefont {M.}~\bibnamefont
  {Toth}},\ }\bibfield  {title} {\enquote {\bibinfo {title} {Charge and energy
  transfer of quantum emitters in 2d heterostructures},}\ }\href {\doibase
  10.1088/2053-1583/ab7fc3} {\bibfield  {journal} {\bibinfo  {journal} {2D
  Materials}\ }\textbf {\bibinfo {volume} {7}},\ \bibinfo {pages} {031001}
  (\bibinfo {year} {2020})}\BibitemShut {NoStop}%
\bibitem [{\citenamefont {Salihoglu}\ \emph {et~al.}(2016)\citenamefont
  {Salihoglu}, \citenamefont {Kakenov}, \citenamefont {Balci}, \citenamefont
  {Balci},\ and\ \citenamefont {Kocabas}}]{Salihoglu_2016}%
  \BibitemOpen
  \bibfield  {author} {\bibinfo {author} {\bibfnamefont {O.}~\bibnamefont
  {Salihoglu}}, \bibinfo {author} {\bibfnamefont {N.}~\bibnamefont {Kakenov}},
  \bibinfo {author} {\bibfnamefont {O.}~\bibnamefont {Balci}}, \bibinfo
  {author} {\bibfnamefont {S.}~\bibnamefont {Balci}}, \ and\ \bibinfo {author}
  {\bibfnamefont {C.}~\bibnamefont {Kocabas}},\ }\bibfield  {title} {\enquote
  {\bibinfo {title} {Graphene as a reversible and spectrally selective
  fluorescence quencher},}\ }\href {\doibase 10.1038/srep33911} {\bibfield
  {journal} {\bibinfo  {journal} {Scientific Reports}\ }\textbf {\bibinfo
  {volume} {6}},\ \bibinfo {pages} {33911} (\bibinfo {year}
  {2016})}\BibitemShut {NoStop}%
\bibitem [{\citenamefont {Lee}\ \emph {et~al.}(2014)\citenamefont {Lee},
  \citenamefont {Bao}, \citenamefont {Ju}, \citenamefont {Schuck},
  \citenamefont {Wang},\ and\ \citenamefont {Weber-Bargioni}}]{Lee_2014}%
  \BibitemOpen
  \bibfield  {author} {\bibinfo {author} {\bibfnamefont {J.}~\bibnamefont
  {Lee}}, \bibinfo {author} {\bibfnamefont {W.}~\bibnamefont {Bao}}, \bibinfo
  {author} {\bibfnamefont {L.}~\bibnamefont {Ju}}, \bibinfo {author}
  {\bibfnamefont {P.~J.}\ \bibnamefont {Schuck}}, \bibinfo {author}
  {\bibfnamefont {F.}~\bibnamefont {Wang}}, \ and\ \bibinfo {author}
  {\bibfnamefont {A.}~\bibnamefont {Weber-Bargioni}},\ }\bibfield  {title}
  {\enquote {\bibinfo {title} {Switching individual quantum dot emission
  through electrically controlling resonant energy transfer to graphene},}\
  }\href {\doibase 10.1021/nl503587z} {\bibfield  {journal} {\bibinfo
  {journal} {Nano Letters}\ }\textbf {\bibinfo {volume} {14}},\ \bibinfo
  {pages} {7115--7119} (\bibinfo {year} {2014})}\BibitemShut {NoStop}%
\bibitem [{\citenamefont {Bjelkevig}\ \emph {et~al.}(2010)\citenamefont
  {Bjelkevig}, \citenamefont {Mi}, \citenamefont {Xiao}, \citenamefont
  {Dowben}, \citenamefont {Wang}, \citenamefont {Mei},\ and\ \citenamefont
  {Kelber}}]{Bjelkevig_2010}%
  \BibitemOpen
  \bibfield  {author} {\bibinfo {author} {\bibfnamefont {C.}~\bibnamefont
  {Bjelkevig}}, \bibinfo {author} {\bibfnamefont {Z.}~\bibnamefont {Mi}},
  \bibinfo {author} {\bibfnamefont {J.}~\bibnamefont {Xiao}}, \bibinfo {author}
  {\bibfnamefont {P.~A.}\ \bibnamefont {Dowben}}, \bibinfo {author}
  {\bibfnamefont {L.}~\bibnamefont {Wang}}, \bibinfo {author} {\bibfnamefont
  {W.-N.}\ \bibnamefont {Mei}}, \ and\ \bibinfo {author} {\bibfnamefont
  {J.~A.}\ \bibnamefont {Kelber}},\ }\bibfield  {title} {\enquote {\bibinfo
  {title} {Electronic structure of a graphene/hexagonal-{BN} heterostructure
  grown on ru(0001) by chemical vapor deposition and atomic layer deposition:
  extrinsically doped graphene},}\ }\href {\doibase
  10.1088/0953-8984/22/30/302002} {\bibfield  {journal} {\bibinfo  {journal}
  {Journal of Physics: Condensed Matter}\ }\textbf {\bibinfo {volume} {22}},\
  \bibinfo {pages} {302002} (\bibinfo {year} {2010})}\BibitemShut {NoStop}%
\bibitem [{Note1()}]{Note1}%
  \BibitemOpen
  \bibinfo {note} {The distributions of the wavefunctions are governed by the
  local external and Hartree potentials (since the orbitals correspond to
  eigenstates of the mean-field Hamiltonian $H_0$).}\BibitemShut {Stop}%
\bibitem [{\citenamefont {Mori-S\'anchez}, \citenamefont {Cohen},\ and\
  \citenamefont {Yang}(2008)}]{Mori_Sanchez_2008}%
  \BibitemOpen
  \bibfield  {author} {\bibinfo {author} {\bibfnamefont {P.}~\bibnamefont
  {Mori-S\'anchez}}, \bibinfo {author} {\bibfnamefont {A.~J.}\ \bibnamefont
  {Cohen}}, \ and\ \bibinfo {author} {\bibfnamefont {W.}~\bibnamefont {Yang}},\
  }\bibfield  {title} {\enquote {\bibinfo {title} {Localization and
  delocalization errors in density functional theory and implications for
  band-gap prediction},}\ }\href {\doibase 10.1103/PhysRevLett.100.146401}
  {\bibfield  {journal} {\bibinfo  {journal} {Phys. Rev. Lett.}\ }\textbf
  {\bibinfo {volume} {100}},\ \bibinfo {pages} {146401} (\bibinfo {year}
  {2008})}\BibitemShut {NoStop}%
\bibitem [{\citenamefont {Cohen}, \citenamefont {Mori-S{\'a}nchez},\ and\
  \citenamefont {Yang}(2008)}]{Cohen_2008}%
  \BibitemOpen
  \bibfield  {author} {\bibinfo {author} {\bibfnamefont {A.~J.}\ \bibnamefont
  {Cohen}}, \bibinfo {author} {\bibfnamefont {P.}~\bibnamefont
  {Mori-S{\'a}nchez}}, \ and\ \bibinfo {author} {\bibfnamefont
  {W.}~\bibnamefont {Yang}},\ }\bibfield  {title} {\enquote {\bibinfo {title}
  {Insights into current limitations of density functional theory},}\ }\href
  {\doibase 10.1126/science.1158722} {\bibfield  {journal} {\bibinfo  {journal}
  {Science}\ }\textbf {\bibinfo {volume} {321}},\ \bibinfo {pages} {792--794}
  (\bibinfo {year} {2008})}\BibitemShut {NoStop}%
\bibitem [{\citenamefont {Kinyanjui}\ \emph {et~al.}(2012)\citenamefont
  {Kinyanjui}, \citenamefont {Kramberger}, \citenamefont {Pichler},
  \citenamefont {Meyer}, \citenamefont {Wachsmuth}, \citenamefont {Benner},\
  and\ \citenamefont {Kaiser}}]{Kinyanjui_2012}%
  \BibitemOpen
  \bibfield  {author} {\bibinfo {author} {\bibfnamefont {M.~K.}\ \bibnamefont
  {Kinyanjui}}, \bibinfo {author} {\bibfnamefont {C.}~\bibnamefont
  {Kramberger}}, \bibinfo {author} {\bibfnamefont {T.}~\bibnamefont {Pichler}},
  \bibinfo {author} {\bibfnamefont {J.~C.}\ \bibnamefont {Meyer}}, \bibinfo
  {author} {\bibfnamefont {P.}~\bibnamefont {Wachsmuth}}, \bibinfo {author}
  {\bibfnamefont {G.}~\bibnamefont {Benner}}, \ and\ \bibinfo {author}
  {\bibfnamefont {U.}~\bibnamefont {Kaiser}},\ }\bibfield  {title} {\enquote
  {\bibinfo {title} {Direct probe of linearly dispersing 2d interband plasmons
  in a free-standing graphene monolayer},}\ }\href {\doibase
  10.1209/0295-5075/97/57005} {\bibfield  {journal} {\bibinfo  {journal} {{EPL}
  (Europhysics Letters)}\ }\textbf {\bibinfo {volume} {97}},\ \bibinfo {pages}
  {57005} (\bibinfo {year} {2012})}\BibitemShut {NoStop}%
\bibitem [{\citenamefont {Towns}\ \emph {et~al.}(2014)\citenamefont {Towns},
  \citenamefont {Cockerill}, \citenamefont {Dahan}, \citenamefont {Foster},
  \citenamefont {Gaither}, \citenamefont {Grimshaw}, \citenamefont {Hazlewood},
  \citenamefont {Lathrop}, \citenamefont {Lifka}, \citenamefont {Peterson},
  \citenamefont {Roskies}, \citenamefont {Scott},\ and\ \citenamefont
  {Wilkins-Diehr}}]{Towns_2014}%
  \BibitemOpen
  \bibfield  {author} {\bibinfo {author} {\bibfnamefont {J.}~\bibnamefont
  {Towns}}, \bibinfo {author} {\bibfnamefont {T.}~\bibnamefont {Cockerill}},
  \bibinfo {author} {\bibfnamefont {M.}~\bibnamefont {Dahan}}, \bibinfo
  {author} {\bibfnamefont {I.}~\bibnamefont {Foster}}, \bibinfo {author}
  {\bibfnamefont {K.}~\bibnamefont {Gaither}}, \bibinfo {author} {\bibfnamefont
  {A.}~\bibnamefont {Grimshaw}}, \bibinfo {author} {\bibfnamefont
  {V.}~\bibnamefont {Hazlewood}}, \bibinfo {author} {\bibfnamefont
  {S.}~\bibnamefont {Lathrop}}, \bibinfo {author} {\bibfnamefont
  {D.}~\bibnamefont {Lifka}}, \bibinfo {author} {\bibfnamefont {G.~D.}\
  \bibnamefont {Peterson}}, \bibinfo {author} {\bibfnamefont {R.}~\bibnamefont
  {Roskies}}, \bibinfo {author} {\bibfnamefont {J.}~\bibnamefont {Scott}}, \
  and\ \bibinfo {author} {\bibfnamefont {N.}~\bibnamefont {Wilkins-Diehr}},\
  }\bibfield  {title} {\enquote {\bibinfo {title} {Xsede: Accelerating
  scientific discovery},}\ }\href {\doibase 10.1109/MCSE.2014.80} {\bibfield
  {journal} {\bibinfo  {journal} {Computing in Science \& Engineering}\ }\textbf
  {\bibinfo {volume} {16}},\ \bibinfo {pages} {62--74} (\bibinfo {year}
  {2014})}\BibitemShut {NoStop}%
\end{thebibliography}%

\end{document}